\documentclass[pra,twocolumn,amsmath,amssymb,showpacs]{revtex4-1}

\usepackage[T1]{fontenc} 
\usepackage{bm}
\usepackage{graphicx}
\usepackage[bookmarks=false]{hyperref}

\newcommand{\bra}[1]{\langle #1|}
\newcommand{\ket}[1]{|#1\rangle}
\newcommand{\braket}[1]{\langle #1 \rangle}
\def\ketg{\ket{\text{g}}}
\def\brag{\bra{\text{g}}}
\def\ketzz{\ket{0,0}}
\def\brazz{\bra{0,0}}

\def\dd{\mathrm{d}}
\def\ee{\mathrm{e}}
\def\ii{\mathrm{i}}
\def\vnabla{\bm{\nabla}}

\def\div{\vnabla\cdot}

\def\Hc{\mathrm{H.c.}}
\def\cc{\mathrm{c.c.}}

\def\kB{\mathrm{k_B}}
\def\PP{\mathrm{P}}

\def\ddt#1{\frac{\partial #1}{\partial t}}

\def\ddz#1{\frac{\partial #1}{\partial z}}

\def\die{\varepsilon}
\def\diez{\varepsilon_0}
\def\diep{\varepsilon_{\text{p}}}
\def\np{n_{\text{p}}}
\def\kp{k_{\text{p}}}
\def\muz{\mu_0}
\def\dm{d}
\def\dens{D}
\def\wa{\omega_{\text{a}}}
\def\wc{\omega_{\text{c}}}
\def\cavlen{\ell}

\def\rabi{g}
\def\rabia{g}
\def\rabiC{\bar{g}}
\def\rabiP{\tilde{g}}
\def\rabiw{\breve{g}}

\def\oH{\hat{\mathcal{H}}}
\def\oHmat{\oH_{\text{mat}}}
\def\oHcav{\oH_{\text{cav}}}
\def\oHenv{\oH_{\text{env}}}
\def\oHenvcav{\oH_{\text{env}}^{\text{cav}}}
\def\oHCO{\oH^{v}_{\infty}}
\def\oHPO{\oH^{r}_{\infty}}
\def\oHCz{\oH^{v}_0}
\def\oHPz{\oH^{r}_0}
\def\oHz{\oH_0}
\def\oHSEC{\oH_{\text{SEC}}}
\def\oHSECcav{\oH_{\text{SEC}}^{\text{cav}}}
\def\oHSECcavRWA{\oH_{\text{SEC}}^{\text{cav-RWA}}}
\def\oHdamp{\oH_{\text{SEC-env}}^{\text{damp}}}
\def\oU{\hat{U}}
\def\oUd{\hat{U}^{\dagger}}

\def\oa{\hat{a}}
\def\oad{\hat{a}^{\dagger}}
\def\osigma{\hat{\sigma}}
\def\osigmad{\hat{\sigma}^{\dagger}}

\def\oLLz{\hat{\mathcal{L}}_0}
\def\oLLother{\hat{\mathcal{L}}_{\text{others}}}
\def\oDDz{\hat{\mathcal{D}}_0}
\def\oDDother{\hat{\mathcal{D}}_{\text{others}}}

\def\oAA{\hat{Q}}
\def\oBB{\hat{\varPi}}
\def\oXX{\hat{X}}
\def\oYY{\hat{Y}}

\def\oA{\hat{A}}
\def\oHf{\hat{H}}

\def\oa{\hat{a}}
\def\oad{\hat{a}^{\dagger}}
\def\oain{\hat{a}_{\text{in}}}
\def\oaind{\hat{a}_{\text{in}}^{\dagger}}
\def\oaout{\hat{a}_{\text{out}}}

\def\op{\hat{p}}
\def\opd{\hat{p}^{\dagger}}

\def\ob{\hat{b}}
\def\obd{\hat{b}^{\dagger}}
\def\obin{\hat{b}^{\text{in}}}

\def\oX{\hat{X}}
\def\oS{\hat{S}}

\def\of{\hat{f}}
\def\ofd{\hat{f}^{\dagger}}
\def\ofc{\hat{f}_{\text{c}}}
\def\ofcd{\hat{f}_{\text{c}}^{\dagger}}
\def\oF{\hat{F}}
\def\oFd{\hat{F}^{\dagger}}
\def\oO{\hat{O}}
\def\orho{\hat{\rho}}
\def\orhoss{\hat{\rho}_{\text{ss}}}

\def\oS{\hat{S}}

\def\oAAp{\hat{Q}^{\downarrow}}
\def\oAAm{\hat{Q}^{\uparrow}}
\def\oXXp{\hat{X}^{\downarrow}}
\def\oXXm{\hat{X}^{\uparrow}}
\def\iAA{\tilde{Q}}
\def\iAAp{\tilde{Q}^{\downarrow}}
\def\iAAm{\tilde{Q}^{\uparrow}}

\def\ovv{\hat{\bm{v}}}
\def\mM{\bm{\mathsf{M}}}
\def\mMk{\bm{C}^{\kappa}}
\def\mMg{\bm{C}^{\gamma}}
\def\munit{\bm{1}}

\def\irho{\tilde{\rho}_{\text{I}}}
\def\irhotot{\tilde{\rho}_{\text{I}}^{\text{tot}}}
\def\irhoB{\tilde{\rho}_{\text{I}}^{\text{B}}}
\def\iHint{\tilde{H}_{\text{SEC}}}
\def\orhotot{\hat{\rho}_{\text{S}}^{\text{tot}}}
\def\oHint{\hat{H}_{\text{SEC}}}
\def\TrB{\mathrm{Tr}_{\text{env}}}

\begin{document}


\title{Maxwell boundary conditions imply non-Lindblad master equation}

\author{Motoaki Bamba}
\altaffiliation{E-mail: bamba@qi.mp.es.osaka-u.ac.jp}
\affiliation{Department of Materials Engineering Science, Osaka University, 1-3 Machikaneyama, Toyonaka, Osaka 560-8531, Japan}
\author{Nobuyuki Imoto}
\affiliation{Department of Materials Engineering Science, Osaka University, 1-3 Machikaneyama, Toyonaka, Osaka 560-8531, Japan}

\date{\today}

\begin{abstract}
From the Hamiltonian connecting the inside and outside of an Fabry-P\'erot cavity,
which is derived from the Maxwell boundary conditions at a mirror of the cavity,
a master equation of a non-Lindblad form is derived
when the cavity embeds matters,
although we can transform it to the Lindblad form by performing the rotating-wave approximation
to the connecting Hamiltonian.
We calculate absorption spectra by these Lindblad and non-Lindblad master equations
and also by the Maxwell boundary conditions in the framework of the classical electrodynamics,
which we consider the most reliable approach.
We found that, compared to the Lindblad master equation,
the absorption spectra by the non-Lindblad one
agree better with those by the Maxwell boundary conditions.
Although the discrepancy is highlighted only in the ultra-strong light-matter interaction regime
with a relatively large broadening,
the master equation of the non-Lindblad form is preferable
rather than of the Lindblad one
for pursuing the consistency with the classical electrodynamics.
\end{abstract}



\maketitle
\section{Introduction}
Most of all realistic systems are not isolated,
but they are coupled with environments.
This system-environment coupling (SEC) is important
for discussing measurements, thermalization, dissipation, noise, etc.
In such open systems,
we in principle require naive consideration on
the separation of systems of interest and environments,
mechanisms of the SECs,
preparation of environment, and so on
\cite{Senitzky1960PR,Dekker1981PR}.
The quantum description of the Brownian motion
has long been discussed as an important subject
of this kind of study
\cite{Caldeira1983PA,Grabert1988PR,Diosi1995EL,Munro1996PRA,gardiner04,Barnett2005PRA,Breuer2006}.
The so-called Caldeira-Leggett master equation \cite{Caldeira1983PA}
is a quantum counterpart of the Fokker-Planck equation
of the Brownian motion.
However, this master equation is not of the Lindblad form
\cite{Lindblad1976CMP}.
Then, the positivity of the density operator is not guaranteed in general.
Whereas we can add extra terms to the Caldeira-Leggett master equation
for transforming it to the Lindblad form, the justification of these extra terms
is unclear in classical physics \cite{Barnett2005PRA,Breuer2006}.
In this way, for the quantum description of open systems,
we sometimes face a trade-off between
the mathematical requirement (positivity of density operator)
and the physical one (consistency with the physical laws).

In this paper, we will show another example for this kind of discussion:
a high-quality Fabry-P\'erot cavity embedding matters with a damping
of excitations.
Such a system has long been discussed in the study of
quantum optics or so-called cavity quantum electrodynamics (cavity QED)
\cite{gardiner04,walls08},
and master equations of the Lindblad form have been used in most cases.
However, we will show that,
from the Hamiltonian of the SEC
derived for the Fabry-P\'erot cavity \cite{Bamba2014SEC},
master equations of a non-Lindblad form are obtained in general.
For transforming it to the Lindblad form,
we need to apply the rotating-wave approximation (RWA)
to the SEC Hamiltonian,
which has been widely used in the study of quantum optics
(but implicitly in most cases).
The influence of the RWA to the SEC was discussed also
in the study of the quantum Brownian motion \cite{Munro1996PRA},
while it was not used in the early study of the quantum description
of dissipation \cite{Senitzky1960PR}.

We will calculate absorption spectra of the cavity system
by the Lindblad and non-Lindblad master equations,
and they will be compared with the results calculated
by the Maxwell boundary conditions (MBCs) at a mirror of the cavity
in the framework of the classical electrodynamics,
which is considered as the most reliable method
(direct consequence by the physical law) in this paper.
Although the quantum theory should reproduce the results in the classical theory,
we sometimes face an inconsistency caused by approximations
used in the quantum theory (RWA to SEC in our case).
We will find that the non-Lindblad master equation gives more consistent
results with those by the MBCs.
The discrepancy between the results by the Lindblad master equation
and by the other two is highlighted
in the ultra-strong light-matter interaction regime
\cite{Ciuti2005PRB} with a relatively rapid damping of excitations in matters.
The ultra-strong interaction means that
the interaction strength is comparable to or larger than the transition frequency of matters,
and it has been realized experimentally in a variety of systems
\cite{Gunter2009N,Anappara2009PRB,Todorov2009PRL,Todorov2010PRL,
Niemczyk2010NP,Fedorov2010PRL,Forn-Diaz2010PRL,
Schwartz2011PRL,Porer2012PRB,Scalari2012S,Zhang2014PRL}.
While both the ultra-strong light-matter interaction
and the rapid damping occur inside the cavity,
the non-Lindblad form concerning the cavity loss,
which is basically supposed much slower than them (good cavity),
will be discussed in this paper.

This paper is organized as follows.
In Sec.~\ref{sec:overview}, we first overview some Lindblad master equations
derived in the past for the cavity system
and the non-Lindblad one to be discussed in this paper.
In Sec.~\ref{sec:master_Langevin}, quantum Langevin equations
that are basically equivalent to the Lindblad and non-Lindblad master equations are shown.
The absorption spectra are in fact calculated by these quantum Langevin equations
for simplifying the calculation.
In Sec.~\ref{sec:model}, we show the models of the cavity and the medium in the cavity.
The Hamiltonians inside the cavity and the SEC are also shown.
In Sec.~\ref{sec:calculation}, we explain the calculation methods
by the MBCs and the quantum Langevin equations.
Typical absorption spectra are also shown in figures.
Sec.~\ref{sec:compare} is devoted to the comparison of the three approaches.
The violation of the positivity in the non-Lindblad master equation
is numerically checked in Sec.~\ref{sec:positivity}.
The advantage of the non-Lindblad master equation
is summarized in Sec.~\ref{sec:advantage},
and the conclusion is shown in Sec.~\ref{sec:summary}.
The detailed derivation of the master and quantum Langevin equations
are shown in App.~\ref{app:derive}.
The master equations for frequency-dependent loss rate are summarized
in App.~\ref{app:master_n}.
The detailed calculation method by the Lindblad-type quantum Langevin equations
is shown in App.~\ref{app:eq_set}.
The absorption spectra calculated by the Lindblad-type treatment
for the excitation damping are discussed in App.~\ref{app:Lindblad_damping}.

\section{Non-Lindblad form to be discussed} \label{sec:overview}
{
\tabcolsep = .5em
\begin{table*}[tbp]
\caption{Validity of three types of master equations
in Eqs.~\eqref{eq:master_a}, \eqref{eq:master_s}, and \eqref{eq:master_non-Lindblad}.
The first two equations are derived under the RWA to the SEC Hamiltonian
but with different bases, and they are of the Lindblad form.
In contrast, Eq.~\eqref{eq:master_non-Lindblad} is derived without the RWA,
and it is of a non-Lindblad form.
The validity of these master equations are summarized
for the weak, normally strong, and ultra-strong light-matter interaction regimes.
They are valid basically for $\omega$-independent loss rate $\kappa$.
The case of $\omega$-dependent $\kappa(\omega)$ is discussed in App.~\ref{app:master_n}
and summarized in Tab.~\ref{tab:A1}.}
\label{tab:1}
\begin{tabular}{lll||lll}
&&& \multicolumn{3}{c}{Light-matter interaction regime} \\
& RWA to SEC && Weak & Strong & Ultra-strong \\ \hline \hline
Eq.~\eqref{eq:master_a} & Photon-based & Lindblad & Good & Good & Bad \\
Eq.~\eqref{eq:master_s} & Eigen-state-based & Lindblad & Good & Good & $^*$Good \\
Eq.~\eqref{eq:master_non-Lindblad} & No & Non-Lindblad & Good & Good & Good \\
\end{tabular}
\\$^*$Good quantitatively for narrow enough broadening avoiding mode overlaps.
\end{table*}
}

In most studies of cavity QED \cite{gardiner04,walls08},
the SEC has been introduced simply
as the injection and escape of photons from cavities
in the framework of master equation, quantum Langevin equation,
stochastic differential equation, or Fokker-Planck equation.
This simple treatment has well reproduced a variety of experimental results.
The SEC Hamiltonian is typically expressed as
\begin{equation} \label{eq:RWA_photon} 
\oH_{\text{SEC}}^{\text{photon}}
= \int_0^{\infty}\dd\omega\
  \ii\hbar\sqrt{\frac{\kappa}{2\pi}} \left[ \ofd(\omega)\oa - \oad\of(\omega) \right].
\end{equation}
Here, $\kappa$ is a frequency-independent loss rate.
$\oa$ is the annihilation operator of a photon in a cavity mode.
$\of(\omega)$ is the annihilation operator of a photon with a frequency $\omega$
outside the cavity, and it satisfies
$[\of(\omega), \ofd(\omega')] = \delta(\omega-\omega')$.
The Hamiltonian of the environment is expressed as
\begin{equation}
\oHenv^{\text{simple}}
= \int_0^{\infty}\dd\omega\ \hbar\omega \ofd(\omega)\of(\omega).
\end{equation}
For the vacuum environment $\braket{\ofd(\omega)\of(\omega')} = 0$
(nearly at zero temperature compared to the frequency scale of interest)
and in the Born-Markov approximation,
the cavity loss is described in the master equation as
\begin{align} \label{eq:master_a} 
\ddt{}\orho
& = \oLLz[\orho]
  + \frac{\kappa}{2} \left(
        \left[ \oa, \orho \oad \right]
      + \left[ \oa\orho, \oad \right]
      \right)
\nonumber \\
& = \oLLz[\orho]
  + \frac{\kappa}{2} \left(
        2\oa\orho \oad - \oad\oa\orho - \orho\oad\oa
      \right).
\end{align}
Here, $\orho$ is the density operator of system of interest (inside the cavity),
and $\oLLz[\orho]$ represents
\begin{equation}
\oLLz[\orho] = \frac{1}{\ii\hbar}\left[\oHz, \orho\right] + \oLLother[\orho],
\end{equation}
where $\oHz$ is the Hamiltonian of the system of interest
and $\oLLother[\orho]$ represents the dissipative terms
originating from couplings with the other environments.

While the master equation such as in Eq.~\eqref{eq:master_a}
has a simple form to be used very easily,
it is in general not appropriate
for strongly coupled composite systems
as discussed in Refs.~\cite{Carmichael1973JPA,Carmichael1974PRA,Scala2007PRA,Scala2007JPA,Fleming2010JPA}.
The system of our interest is also such a composite system, i.e.,
photons and excitations in matters.
Instead of Eq.~\eqref{eq:master_a},
the master equation should be derived under specifying
a (non-dimensional) physical quantity $\oAA$ of the system of interest
that mediates the SEC.
From the detail of the mechanism of SEC,
the SEC Hamiltonian is typically derived as \cite{Bamba2014SEC}
\begin{equation} \label{eq:oHSEC_simple} 
\oHSEC^{\text{simple}}
= \int_0^{\infty}\dd\omega\
  \ii\hbar\sqrt{\frac{\kappa}{2\pi}} \oAA \left[ \ofd(\omega) - \of(\omega) \right].
\end{equation}
Here, applying the RWA to
this SEC Hamiltonian in the basis of the eigen-states $\{\ket{\mu}\}$ of $\oHz$, we get
\begin{equation} \label{eq:RWA_eigen} 
\oHSEC^{\text{simple}}
\approx \int_0^{\infty}\dd\omega\
  \ii\hbar\sqrt{\frac{\kappa}{2\pi}} \left[ \ofd(\omega)\oAAp - \oAAm\of(\omega) \right],
\end{equation}
where $\oAAp$ and $\oAAm$ are the lowering and raising components of $\oAA$,
respectively, as
\begin{subequations} \label{eq:lower_raise} 
\begin{align}
\oAAp & \equiv \sum_{\mu}\sum_{\nu>\mu} \oAA_{\mu,\nu}, \\
\oAAm & \equiv \sum_{\mu}\sum_{\nu>\mu} \{\oAA_{\mu,\nu}\}^{\dagger}
= \{\oAAp\}^{\dagger},
\end{align}
\end{subequations}
\begin{equation}
\oAA_{\mu,\nu} \equiv \ket{\mu}\braket{\mu|\oAA|\nu}\bra{\nu}
= \{\oAA_{\nu,\mu}\}^{\dagger}.
\end{equation}
Note that, in this paper, operators mediating SECs are supposed to have no diagonal matrix
element $\braket{\mu|\oAA|\mu} = 0$ in the basis of the eigen-states,
whereas such a diagonal element causes the pure dephasing in principle
\cite{walls08,Beaudoin2011PRA}.
From the approximated SEC Hamiltonian in Eq.~\eqref{eq:RWA_eigen},
the master equation is derived in the Born-Markov approximation as
\begin{align} \label{eq:master_s} 
\ddt{}\orho
& = \oLLz[\orho]
  + \frac{\kappa}{2} \left(
        \left[ \oAAp, \orho \oAAm \right]
      + \left[ \oAAp\orho, \oAAm \right]
      \right)
\nonumber \\
& = \oLLz[\orho]
  + \frac{\kappa}{2} \left(
        2\oAAp\orho \oAAm - \oAAm\oAAp\orho - \orho\oAAm\oAAp
      \right).
\end{align}
In contrast, if the interaction in the composite system
is weak enough compared with the loss rate $\kappa$
(weak interaction regime),
the RWA to the SEC in the photon basis can be justified,
and the widely-used SEC Hamiltonian in Eq.~\eqref{eq:RWA_photon}
is derived by supposing $\oAA = \oa + \oad$
or $\oAA = \ii(\oa-\oad)$.
Then, the simple master equation in Eq.~\eqref{eq:master_a} is derived
in the Born-Markov approximation.

The master equation in Eq.~\eqref{eq:master_s}
is not equivalent to the simple one in Eq.~\eqref{eq:master_a}
if the light-matter interaction is in the ultra-strong regime \cite{Beaudoin2011PRA}.
For example, here we tentatively suppose a simple Hamiltonian
\begin{equation} \label{eq:oHz_simple} 
\oHz^{\text{simple}} = \hbar\wc\oad\oa + \hbar\rabiw(\oa+\oad)\osigma_x + \oHmat.
\end{equation}
Here, $\wc$ is the resonance frequency of a cavity mode.
$\oHmat$ is the Hamiltonian of matters inside the cavity,
and $\osigma_x = \osigma + \osigmad$ is a non-dimensional operator
that annihilates ($\osigma$) or creates ($\osigmad$) an excitation in matters.
$\rabiw$ represents the strength of the light-matter interaction,
and the ultra-strong regime means 
$\rabiw \gtrsim \wc, \wa$,
where $\wa$ is a characteristic transition frequency of the matter.
Due to the so-called counter-rotating terms $\oa\osigma$ and $\oad\osigmad$
in $\oHz^{\text{simple}}$,
the annihilation of a photon $\oa$ no longer
corresponds to the lowering operator
for the system inside the cavity ($\oa \neq \oAAp$) \cite{Ciuti2005PRB}.
Then, in the ultra-strong interaction regime,
we must use the master equation in Eq.~\eqref{eq:master_s} \cite{Beaudoin2011PRA}.
If we use the simple master equation in Eq.~\eqref{eq:master_a}
for the Hamiltonian with the counter-rotating terms in Eq.~\eqref{eq:oHz_simple},
the system inside the cavity is in general excited even by the vacuum environment
\cite{Bamba2012DissipationUSC}.
The degree of the excitation can no longer be negligible
in the ultra-strong interaction regime.

In contrast, in the normally strong interaction regime
($\kappa \ll \rabiw \ll \wc,\wa$),
we can apply the RWA to the light-matter interaction,
and the counter-rotating terms are eliminated as
\begin{equation}
\oHz^{\text{simple}} \approx \hbar\wc\oad\oa + \hbar\rabiw(\osigmad\oa+\oad\osigma) + \oHmat.
\end{equation}
For this approximated Hamiltonian,
the photon annihilation corresponds to the lowering of the cavity system
($\oa = \oAAp$).
Then, the master equation in Eq.~\eqref{eq:master_s}
is in fact reduced to the simple one in Eq.~\eqref{eq:master_a}.
In Tab.~\ref{tab:1},
we summarize the validity of the two master equations in 
Eqs.~\eqref{eq:master_a} and \eqref{eq:master_s}.

Note that the simple master equation in Eq.~\eqref{eq:master_a}
is valid in the normally strong interaction regime
only for the $\omega$-independent loss rate $\kappa$.
If the loss rate $\kappa(\omega)$ relatively varies
in the frequency range of interest,
we need to use the extended version of the master equation in Eq.~\eqref{eq:master_s}
applicable to the $\omega$-dependent $\kappa(\omega)$
(see the detail in App.~\ref{app:master_n}).
Such a master equation can no longer be reduced to the simple one
as in Eq.~\eqref{eq:master_a}.
Further, the extended master equation is of a non-Lindblad form,
while both Eqs.~\eqref{eq:master_a} and \eqref{eq:master_s}
for $\omega$-independent $\kappa$ are of the Lindblad form.
In this way, when we consider the $\omega$-dependent loss rate $\kappa(\omega)$,
we have faced the problem of the Lindblad form
(positivity of density operator)
in the study of cavity QED
even in the weak and normally strong interaction regimes.
The detail is discussed in App.~\ref{app:master_n}.

However, the non-Lindblad form
caused by the $\omega$-dependent $\kappa(\omega)$ is not the target in this paper.
We will argue that the following non-Lindblad master equation
should be used even for the $\omega$-independent $\kappa$:
\begin{align} \label{eq:master_non-Lindblad} 
\ddt{}\orho
& = \oLLz[\orho]
  + \frac{\kappa}{2} \left(
        \left[ \oAA, \orho \oAAm \right]
      + \left[ \oAAp\orho, \oAA \right]
      \right).
\end{align}
This is derived also in the Born-Markov approximation.
But, in contrast to Eq.~\eqref{eq:master_s},
this is derived directly from the SEC Hamiltonian in Eq.~\eqref{eq:oHSEC_simple}
without the RWA performed in Eq.~\eqref{eq:RWA_eigen}
(the detailed derivation is shown in App.~\ref{app:derive}).
For discussing the necessity of this non-Lindblad form in Eq.~\eqref{eq:master_non-Lindblad},
we will basically suppose $\omega$-independent $\kappa$ in the following sections.
Whereas the Lindblad master equation in Eq.~\eqref{eq:master_s}
was recognized to be applicable basically in the ultra-strong interaction regime,
we will show that the absorption spectra by it in general deviate from
those by the MBCs,
which are supposed to be the most reliable approach in this paper.
We get larger deviation for stronger light-matter interaction
and wider broadening (this is the reason of $^*$Good in Tab.~\ref{tab:1}).
We will show that
the non-Lindblad master equation in Eq.~\eqref{eq:master_non-Lindblad}
gives more consistent results with those by the MBCs
than the Lindblad one in Eq.~\eqref{eq:master_s}.

Note that,
by extending the master equations in Eqs.~\eqref{eq:master_s} and \eqref{eq:master_non-Lindblad}
for the environments at a non-zero temperature $T$,
we obtain the thermal state $\rho = \ee^{-\oHz/\kB T}$
as a steady state of such extended master equations
of both the Lindblad and non-Lindblad forms.
The detail is discussed in App.~\ref{app:master_n}.

Note also that we basically neglect the energy shift (Lamb shift) due to the SEC
both in the Lindblad and non-Lindblad master equations.
While the energy shift is implicitly included
in the calculation of absorption spectra by the MBCs,
it will not clearly appear
in the broad absorption peaks in our numerical calculations.

\section{Master and Langevin equations} \label{sec:master_Langevin}
Whereas we aim to discuss the validity of
the Lindblad and non-Lindblad master equations
as shown in Eqs.~\eqref{eq:master_s} and \eqref{eq:master_non-Lindblad},
we will in fact calculate absorption spectra by quantum Langevin equations
corresponding to those two master equations
for pursuing simple and clear calculations.
In this section, we preliminarily shows this correspondence.

We will consider the SEC Hamiltonian expressed as
\begin{equation} \label{eq:oHSEC_cavity} 
\oHSECcav
= \int_0^{\infty}\dd\omega
  \sum_j\ii\hbar\sqrt{\frac{\kappa_j}{2\pi}} \oAA_j \left[ \ofcd(\omega) - \ofc(\omega) \right].
\end{equation}
Basically, all the modes indexed by $j$ in a Fabry-P\'erot cavity
are considered, and
\begin{equation}
\oAA_j = \oa_j + \oad_j
\end{equation}
is a coordinate of the $j$-th mode.
$\oa_j$ is the annihilation operator of a photon,
and $\kappa_j$ is the loss rate of this mode.
All the cavity modes couple with the same environment
described by $\ofc(\omega)$, whose correlation is supposed as
\begin{subequations}
\begin{align}
\braket{\ofcd(\omega)\ofc(\omega')} & = 0, \\
\braket{\ofc(\omega)\ofcd(\omega')} & = \delta(\omega-\omega').
\end{align}
\end{subequations}
The Hamiltonian of the environment is expressed as
\begin{equation}
\oHenvcav
= \int_0^{\infty}\dd\omega\ \hbar\omega\ofcd(\omega)\ofc(\omega).
\end{equation}
Applying the RWA to Eq.~\eqref{eq:oHSEC_cavity}, we get
\begin{equation} \label{eq:oHSEC_cavity_RWA} 
\oHSECcavRWA
= \int_0^{\infty}\dd\omega\
  \sum_j \ii\hbar\sqrt{\frac{\kappa_j}{2\pi}}
  \left[ \ofcd(\omega)\oAAp_j - \oAAm_j\ofc(\omega) \right],
\end{equation}
where $\oAAp_j$ and $\oAAm_j$ are the lowering
and raising components of $\oAA_j$, respectively,
similarly defined as in Eqs.~\eqref{eq:lower_raise}.
From Eqs.~\eqref{eq:oHSEC_cavity} and \eqref{eq:oHSEC_cavity_RWA},
the non-Lindblad and Lindblad master equations are derived,
respectively, as
\begin{subequations}
\begin{align}
\ddt{}\orho
& = \oLLz[\orho]
  + \sum_{j,j'}
    \frac{\sqrt{\kappa_j\kappa_{j'}}}{2}
      \left(
        \left[ \oAA_j, \orho \oAAm_{j'} \right]
      + \left[ \oAAp_{j'}\orho, \oAA_j \right]
      \right), \label{eq:master_non-Lindblad_jj} \\ 
\ddt{}\orho
& = \oLLz[\orho]
  + \sum_{j,j'}
    \frac{\sqrt{\kappa_j\kappa_{j'}}}{2}
      \left(
        \left[ \oAAp_j, \orho \oAAm_{j'} \right]
      + \left[ \oAAp_{j'}\orho, \oAAm_j \right]
      \right). \label{eq:master_Lindblad_jj} 
\end{align}
\end{subequations}
The derivation of them are shown in App.~\ref{app:derive}.

On the other hand, 
from Eqs.~\eqref{eq:oHSEC_cavity} and \eqref{eq:oHSEC_cavity_RWA},
quantum Langevin equations for arbitrary system operator $\oS$ are derived
(detail is shown also in App.~\ref{app:derive}), respectively, as
\begin{subequations}
\begin{align}
\ddt{}\oS(t)
& = \oDDz[\oS]
- \int_0^{\infty}\dd\omega\ \left\{
  \sum_j \left[ \oS, \oAA_j \right] \ee^{-\ii\omega t}\sqrt{\kappa_j}
\right. \nonumber \\ & \quad \left. \times
  \left[ \oain(\omega) + \sum_{j'}\frac{\sqrt{\kappa_{j'}}}{2}\oAA_{j'}(\omega) \right]
  + \Hc
  \right\}, \label{eq:Langevin_non-Lindblad} \\ 
\ddt{}\oS(t)
& = \oDDz[\oS]
- \int_0^{\infty}\dd\omega\ \left\{
  \sum_j \left[ \oS, \oAAm_j \right] \ee^{-\ii\omega t}\sqrt{\kappa_j}
\right. \nonumber \\ & \quad \left. \times
  \left[ \oain(\omega) + \sum_{j'}\frac{\sqrt{\kappa_{j'}}}{2}\oAAp_{j'}(\omega) \right]
  + \Hc
  \right\}, \label{eq:Langevin_Lindblad} 
\end{align}
\end{subequations}
where the first terms are represented as
\begin{equation}
\oDDz[\oS]
= \frac{1}{\ii\hbar}\left[ \oS, \oHz \right]
+ \oDDother[\oO].
\end{equation}
$\oDDother[\oS]$ includes the dissipation and noise terms by the other environment.
The Fourier transform of operators in the Heisenberg picture is defined as
\begin{equation}
\oAA_j(\omega) = \frac{1}{2\pi}\int_{-\infty}^{\infty}\dd t\ \ee^{\ii\omega t} \oAA_j(t).
\end{equation}
$\oain(\omega)$ is the so-called input operator \cite{Blow1990PRA,walls08,gardiner04}
satisfying
\begin{equation}
[ \oain(\omega), \oaind(\omega')] = \delta(\omega-\omega')/2\pi.
\end{equation}
From the SEC Hamiltonians in Eqs.~\eqref{eq:oHSEC_cavity} and \eqref{eq:oHSEC_cavity_RWA},
the input-output relations are derived, respectively, as
\begin{subequations} \label{eq:input-output} 
\begin{align}
\oaout(\omega) & = \oain(\omega) + \sum_j \sqrt{\kappa_j}\oAA_j(\omega),
\label{eq:inout_non-Lindblad} \\
\oaout(\omega) & = \oain(\omega) + \sum_j \sqrt{\kappa_j}\oAAp_j(\omega).
\label{eq:inout_Lindblad}
\end{align}
\end{subequations}

The quantum Langevin equation in Eq.~\eqref{eq:Langevin_non-Lindblad} corresponds to the non-Lindblad master equation
in Eq.~\eqref{eq:master_non-Lindblad_jj},
because they are derived from the same SEC Hamiltonian
in Eq.~\eqref{eq:oHSEC_cavity}.
On the other hand, 
Eq.~\eqref{eq:Langevin_Lindblad} corresponds to the Lindblad one
in Eq.~\eqref{eq:master_Lindblad_jj},
which are derived from Eq.~\eqref{eq:oHSEC_cavity_RWA}.
For checking the correspondence,
let us derive equations of motion of expectation values.
The equations of motion of $\braket{\oS(t)}$ are derived from
the master equations in
Eqs.~\eqref{eq:master_non-Lindblad_jj} and Eq.~\eqref{eq:master_Lindblad_jj},
respectively, as
\begin{subequations}
\begin{align}
\ddt{}\braket{\oS(t)}
& = \frac{1}{\ii\hbar}\braket{[\oS, \oHz]} + \ldots
\nonumber \\ & \quad
  - \sum_{j,j'}
    \frac{\sqrt{\kappa_j\kappa_{j'}}}{2}
      \left\{
        \braket{ [ \oS, \oAA_j ] \oAAp_{j'}(t)}
        + \cc
      \right\}, \label{eq:exp1a} \\
\ddt{}\braket{\oS(t)}
& = \frac{1}{\ii\hbar}\braket{[\oS, \oHz]} + \ldots
\nonumber \\ & \quad
  - \sum_{j,j'}
    \frac{\sqrt{\kappa_j\kappa_{j'}}}{2}
      \left\{
        \braket{ [ \oS, \oAAm_j ] \oAAp_{j'}(t)}
        + \cc
      \right\}. \label{eq:exp1b}
\end{align}
\end{subequations}
On the other hand, from the quantum Langevin equations
in Eq.~\eqref{eq:Langevin_non-Lindblad} and Eq.~\eqref{eq:Langevin_Lindblad},
respectively, we get
\begin{subequations}
\begin{align}
\ddt{}\braket{\oS(t)}
& = \frac{1}{\ii\hbar}\braket{[\oS, \oHz]} + \ldots
- \sum_{j,j'}   \frac{\sqrt{\kappa_j\kappa_{j'}}}{2}
\nonumber \\ & \quad \times
  \int_0^{\infty}\dd\omega \left\{
  \ee^{-\ii\omega t}
  \braket{[\oS,\oAA_j]\oAA_{j'}(\omega)}
  + \cc
  \right\}, \label{eq:exp2a} \\
\ddt{}\braket{\oS(t)}
& = \frac{1}{\ii\hbar}\braket{[\oS, \oHz]} + \ldots
- \sum_{j,j'}   \frac{\sqrt{\kappa_j\kappa_{j'}}}{2}
\nonumber \\ & \quad \times
  \int_0^{\infty}\dd\omega \left\{
  \ee^{-\ii\omega t}
  \braket{[\oS,\oAAm_j]\oAAp_{j'}(\omega)}
  + \cc
  \right\}. \label{eq:exp2b}
\end{align}
\end{subequations}
In the absence of the SEC, the lowering component corresponds exactly to
the positive-frequency component as
\begin{equation}
\oAAp_j(t) = \int_0^{\infty}\dd\omega\ \ee^{-\ii\omega t}\oAA_j(\omega),
\end{equation}
\begin{equation}
\oAAp_j(\omega)
\equiv \frac{1}{2\pi}\int_{-\infty}^{\infty}\dd t\ \ee^{\ii\omega t}\oAAp_j(t)
= \oAA_j(\omega)
\quad \text{for $\omega > 0$}.
\end{equation}
Even in the presence of the SEC,
this equivalence is approximately guaranteed,
if the line broadening does not elongate around $\omega = 0$.
We basically suppose such a situation in the main text of this paper,
and the dissipative terms approximately equal
between Eq.~\eqref{eq:exp1a} and Eq.~\eqref{eq:exp2a}
and between Eq.~\eqref{eq:exp1b} and Eq.~\eqref{eq:exp2b}.
In fact, what is more important is the commutators:
$[\oS, \oAA_j]$ for the non-Lindblad form
and $[\oS, \oAAm_j]$ for the Lindblad form.
In this way,
the quantum Langevin equations in Eq.~\eqref{eq:Langevin_non-Lindblad}
and \eqref{eq:Langevin_Lindblad}
correspond to the non-Lindblad master equation
in Eq.~\eqref{eq:master_non-Lindblad_jj}
and the Lindblad one in Eq.~\eqref{eq:master_Lindblad_jj},
respectively.
We will calculate absorption spectra by these quantum Langevin equations.

\section{Model and Hamiltonian} \label{sec:model}
\begin{figure}[tbp]
\begin{center}
\includegraphics[width=.9\linewidth]{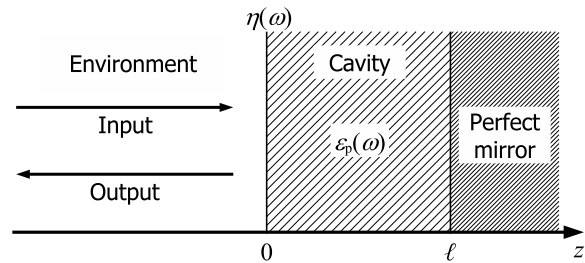}
\end{center}
\caption{The Fabry-P\'erot cavity considered in this paper.
A perfect and imperfect mirrors are placed at $z = \cavlen$ and 0, respectively.
The inside of the cavity is filled by a medium with a dielectric function $\diep(\omega)$.}
\label{fig:FPcavity}
\end{figure}
In order to evaluate the validity of the Lindblad and non-Lindblad master equations
shown in the previous sections,
we consider a Fabry-P\'erot cavity embedding a dispersive and absorptive
medium depicted in Fig.~\ref{fig:FPcavity}.
We will compare absorption spectra calculated by these master equations
(quantum Langevin equations exactly speaking)
and by the MBCs,
which are shown in Sec.~\ref{sec:MBCs}.
Perfect and imperfect mirrors are placed at $z = \cavlen$ and 0, respectively.
Supposing the spatially dependent dielectric function $\die(z,\omega)$ of this cavity structure,
we derive a SEC Hamiltonian connecting inside and outside the cavity
in Sec.~\ref{sec:oHSEC}.
The Hamiltonian describing the cavity modes and the dispersive
and absorptive medium is shown in Sec.~\ref{sec:oHz}.
From these Hamiltonians,
we will calculate absorption spectra
by the Lindblad- and non-Lindblad-type quantum Langevin equations
in Sec.~\ref{sec:compare}.
On the other hand, in Sec.~\ref{sec:diep},
from the Hamiltonian inside the cavity,
we calculate a dielectric function $\diep(\omega)$ (dispersion relation) of the electromagnetic wave in the medium.
From this dielectric function and that of the mirrors,
we will also calculate absorption spectra by the MBCs
in Secs.~\ref{sec:calculation} and \ref{sec:compare}.
We will compare these three absorption spectra
for evaluating the validity of the Lindblad and non-Lindblad master equations
in Sec.~\ref{sec:compare}.
In the following sub-sections, we explain the model of the cavity system
and the Hamiltonian describing it.

\subsection{Maxwell boundary conditions} \label{sec:MBCs}
In the same manner as in Refs.~\cite{Lang1973PRA,Bamba2013MBC,Bamba2014SEC},
we describe the imperfect mirror by the Kronecker's delta function
with a coefficient $\eta(\omega)$,
and the dielectric function of the system is expressed as
\begin{equation} \label{eq:die(z,w)} 
\die(z,\omega) = \eta(\omega) \delta(z)
+ \begin{cases}
0 & z < 0 \\
\diep(\omega) & 0 < z < \cavlen \\
\infty & \cavlen < z
\end{cases}
\end{equation}
The Maxwell boundary conditions at $z = 0$ are derived
for the electric field $E(z,\omega)$ and the magnetic one $H(z,\omega)$ as
\cite{Bamba2013MBC,Bamba2014SEC}
\begin{subequations} \label{eq:MBC} 
\begin{align}
E(0^-,\omega) & = E(0^+,\omega), \\
H(0^-,\omega) - H(0^+,\omega) & = -\ii\omega\diez\eta(\omega) E(0,\omega).
\end{align}
\end{subequations}
These are independent of the detail of $\diep(\omega)$,
i.e., the detail inside the cavity.
Further, at $z = \cavlen$, the electric field satisfies
\begin{equation}
E(\cavlen,\omega) = 0.
\end{equation}
The electric and magnetic fields are expressed by the vector potential $A(z,\omega)$ as
\begin{subequations}
\begin{align}
E(z,\omega) & = \ii\omega A(z,\omega), \\
H(z,\omega) & = \frac{1}{\muz}\ddz{}A(z,\omega)
= \frac{1}{\ii\omega\muz}\ddz{}E(z,\omega).
\end{align}
\end{subequations}

\subsection{Hamiltonian of system-environment coupling} \label{sec:oHSEC}
From the above Maxwell boundary conditions,
in Ref.~\cite{Bamba2014SEC},
we derived the Hamiltonian connecting inside and outside the cavity,
which is independent of the detail inside the cavity
but is valid only for good cavities.
In the absence of the SEC (in the case of perfect cavity)
and of the light-matter interaction,
the Hamiltonian of the electromagnetic fields
inside the cavity is simply expressed as
\begin{equation} \label{eq:oHcav} 
\oHcav = \sum_j \hbar c k_j \oad_j \oa_j,
\end{equation}
where $\oa_j$ annihilates a photon in the $j$-th mode 
satisfying $[ \oa_j, \oad_{j'} ] = \delta_{j,j'}$ and 
\begin{equation} \label{eq:k_j} 
k_j = \frac{j\pi}{\cavlen},\quad j = 1, 2, 3, \ldots
\end{equation}
is the confinement wavenumber.
The operators of the vector potential and magnetic field
inside the cavity are expressed as
\begin{subequations}
\begin{align}
\oA(z)
& = \sum_{j} \sqrt{\frac{\hbar}{\diez ck_j\cavlen}} \sin(k_jz)
    \left( \oa_j + \oad_j \right), \\
\oHf(z)
& = \sum_{j} \frac{k_j}{\muz}\sqrt{\frac{\hbar}{\diez ck_j\cavlen}} \cos(k_jz)
    \left( \oa_j + \oad_j \right).
\end{align}
\end{subequations}
Introducing a non-dimensional quantity
\begin{equation}
\varLambda(\omega) = \eta(\omega)\omega/c
\end{equation}
($\varLambda \gg 1$ corresponds to good cavity),
the SEC Hamiltonian is described
by the magnetic field $\oHf(0^+)$ at the imperfect mirror $z = 0$ as
\cite{Bamba2014SEC}
\begin{align} \label{eq:oHSECcav} 
\oHSECcav
& = \int_0^{\infty}\dd\omega\
    \ii\hbar\sqrt{\frac{\muz c}{\pi\hbar\omega\varLambda(\omega)^2}}
    \left[ \ofcd(\omega) - \ofc(\omega) \right]
    \oHf(0^+) \nonumber \\
& = \sum_j \int_0^{\infty}\dd\omega\
    \ii\hbar\sqrt{\frac{\kappa_j(\omega)}{2\pi}}
    \left[ \ofcd(\omega) - \ofc(\omega) \right]
    \left( \oa_j + \oad_j \right),
\end{align}
where the loss rate $\kappa_j(\omega)$ for empty cavity is expressed as
\begin{equation} \label{eq:kappa_j} 
\kappa_j(\omega) = \frac{2c^2k_j}{\omega\cavlen\varLambda(\omega)^2}.
\end{equation}
From the MBCs in Eqs.~\eqref{eq:MBC},
the reflectance of the imperfect mirror is obtained as
\begin{equation}
R_{\text{mirror}}(\omega) = \frac{1}{1 + 4/\varLambda(\omega)^2}.
\end{equation}
Since the round trip time of light inside the cavity is $2\cavlen / c$,
the loss rate $\kappa_j$ (decay rate of light intensity)
of the $j$-th mode is estimated as
\begin{equation}
K_j
= \frac{1-R_{\text{mirror}}(\omega_j)}{2\cavlen/c}
= \frac{c}{2\cavlen}\frac{1}{1+\varLambda(\omega_j)^2/4}.
\end{equation}
In the good cavity limit $\varLambda(\omega) \gg 1$,
$\kappa_j(\omega_j)$ in Eq.~\eqref{eq:kappa_j} is certainly equal to $K_j$.

Since we will consider basically the $\omega$-independent $\kappa$,
we suppose $\eta(\omega) \propto \omega^{-3/2}$
and
\begin{equation}
\varLambda(\omega) = \varLambda_0\sqrt{\frac{\wa}{\omega}}.
\end{equation}
Here, $\varLambda_0$ is independent of $\omega$,
and we will basically suppose $\varLambda_0 = 10^3 \gg 1$
in the numerical calculation.

\subsection{Dielectric function of medium} \label{sec:diep}
As the model of the dispersive and absorptive medium inside the cavity,
we consider a bosonic excitation with a transition frequency $\wa$,
transition dipole moment $\dm$,
density $\dens$ of excitonic sites,
and infinite translational mass of excitation.
Here, we first consider a spatially infinite medium,
and derive its dielectric function $\diep(\omega)$.
As discussed, for example, in Refs.~\cite{cohen-tannoudji89,Bamba2014SPT},
when we restrict the light propagation in the $z$ direction,
the Hamiltonian of such a medium can be described in the velocity or length form
(sometimes called the Coulomb and electric dipole gauges
while both are in the Coulomb gauge in the sense of $\div\bm{A} = 0$)
equivalently as
\begin{subequations} \label{eq:oH1D} 
\begin{align}
\oHCO/\hbar
& = \sum_{k} \left[
    c|k| \oad_{k}\oa_{k}
  + \wa \obd_k \ob_k
  + \rabiC_{k} \oAA_k \oYY_{-k}
\right. \nonumber \\ & \quad \left.
  + (\rabiC_k{}^2/\wa) \oAA_k \oAA_{-k}
    \right], \label{eq:oHCO} \\ 
\oHPO/\hbar
& = \sum_{k} \left[
    c|k| \oad_{k}\oa_{k}
  + \wa \obd_k \ob_k
  - \rabiP_{k} \oBB_k \oXX_{-k}
\right. \nonumber \\ & \quad \left.
  + (\rabiP_{k}{}^2/c|k|) \oXX_{k} \oXX_{-k}
  \right].
\label{eq:oHPO} 
\end{align}
\end{subequations}
Here, $\oa_k$ and $\ob_k$ are annihilation operators of a photon
and a bosonic excitation with a wavenumber $k$.
They satisfy
$[\oa_k, \oad_{k'}] = [\ob_k, \obd_{k'}] = \delta_{k,k'}$,
and the other combinations are commutable.
The capital operators are Hermitian and defined as
\begin{subequations} \label{eq:oABXY} 
\begin{align}
\oAA_{k} & = \oa_k + \oad_{-k}, \\
\oBB_{k} & = \ii\left(\oa_k - \oad_{-k} \right), \\
\oXX_{k} & = \ob_k + \obd_{-k}, \\
\oYY_{k} & = \ii\left(\ob_k - \obd_{-k} \right).
\end{align}
\end{subequations}
They satisfy
$[\oAA_k,\oBB_{k'}] = [\oXX_k,\oYY_{k'}] = -\ii2\delta_{k,k'}$,
and the other combinations are commutable.
The light-matter interaction strengths are expressed
for a non-dimensional strength $\rabia$ as
\begin{equation} \label{eq:rabi} 
\rabiC_k
= \rabia \wa \sqrt{\frac{\wa}{c|k|}}, \quad
\rabiP_k
= \rabia c|k| \sqrt{\frac{\wa}{c|k|}}, \quad
\rabia
= \sqrt{\frac{\dens|\dm|^2}{2\diez\hbar\wa}}.
\end{equation}
The ultra-strong interaction means $\rabi \gtrsim 1$ in this paper.
Using an unitary operator
\begin{equation} \label{eq:oU} 
\oU = \exp\left[ - \ii\rabia \sum_k \sqrt{\frac{\wa}{c|k|}} \oAA_k\oXX_{-k} \right],
\end{equation}
the two Hamiltonians in Eqs.~\eqref{eq:oH1D} are transformed to each other as
\cite{cohen-tannoudji89,Bamba2014SPT}
\begin{equation}
\oU\oHCO\oUd = \oHPO.
\end{equation}

Since we will calculate the absorption (reflection) spectra
of the cavity system for evaluating the validity
of the Lindblad and non-Lindblad master equations,
we consider also a damping of excitations in matters.
Otherwise, the absorption is never observed.
For keeping the equivalence between the velocity and length forms,
we consider that the damping is mediated by $\{\oXX_k\}$,
which is commutable with the unitary operator $\oU$ in Eq.~\eqref{eq:oU}.
The environment for the damping is introduced for each excitation mode independently,
and the Hamiltonian of the environment and the SEC for the damping
is described as
\begin{align}
\oH_{\infty}^{\text{damp}}
& = \sum_k\int_0^{\infty}\dd\omega\left\{
    \hbar\omega\ofd_k(\omega)\of_k(\omega)
\right. \nonumber \\ & \quad \left.
  + \ii\hbar\sqrt{\frac{\gamma}{2\pi}}
    \left[ \ofd_k(\omega) - \of_k(\omega) \right] \oX_k
  \right\}.
\end{align}
Here, $\of_k(\omega)$ is the annihilation operator of a boson
in the environment for mode $k$,
and it satisfies
$[\of_k(\omega), \ofd_{k'}(\omega')] = \delta_{k,k'}\delta(\omega-\omega')$
and $\braket{\ofd_k(\omega)\of_{k'}(\omega')} = 0$.
We suppose that the damping rate $\gamma$ is $\omega$-independent
for discussing the non-Lindblad form focused in this paper.
For deriving the dielectric function $\diep(\omega)$ of this damping medium
(dispersive and absorptive medium),
in the same manner as in Eq.~\eqref{eq:Langevin_non-Lindblad},
we consider the following quantum Langevin equation
corresponding to the non-Lindblad master equation:
\begin{align}
\ddt{}\oS
& = \frac{1}{\ii\hbar}\left[ \oS, \oH_0 \right]
- \sum_k \int_0^{\infty}\dd\omega\ \left\{
  \left[ \oS, \oXX_k \right] \ee^{-\ii\omega t}
\right. \nonumber \\ & \quad \left. \times
  \left[\frac{\gamma}{2}\oXX_k(\omega) + \sqrt{\gamma}\obin_k(\omega)\right]
  + \Hc
  \right\},
\end{align}
where $\obin_k$ is the input operator due to the damping.
The discussion for Lindblad-type damping
is performed in App.~\ref{app:Lindblad_damping}.
Since there is no loss of photons in the infinite medium,
the quantum Langevin equations are derived
in the velocity form as
\begin{subequations} \label{eq:motion_Coulomb} 
\begin{align}
-\ii\omega\oAA_{k}(\omega) & = - c|k| \oBB_{k}(\omega), \\
-\ii\omega\oBB_{k}(\omega) & = (c|k|+4\rabiC_{k}{}^2/\wa) \oAA_{k}(\omega) + 2\rabiC_{k}\oYY_{k}(\omega), \\
-\ii\omega\oXX_{k}(\omega) & = -\wa\oYY_{k}(\omega) - 2\rabiC_{k}\oAA_{k}(\omega), \\
-\ii\omega\oYY_{k}(\omega) & = (\wa-\ii\gamma) \oXX_{k}(\omega) - \ii2\sqrt{\gamma}\obin_k(\omega).
\end{align}
\end{subequations}
On the other hand, in the length form, we get
\begin{subequations}
\begin{align}
-\ii\omega\oAA_{k}(\omega)
& = - c|k| \oBB_{k}(\omega) + 2\rabiP_{k} \oXX_{k}(\omega), \\
-\ii\omega\oBB_{k}(\omega)
& = c|k| \oAA_{k}(\omega), \\
-\ii\omega\oXX_{k}(\omega)
& = - \wa \oYY_{k}(\omega), \\
-\ii\omega\oYY_{k}(\omega)
& = (\wa-\ii\gamma) \oXX_{k}(\omega)  - \ii2\sqrt{\gamma}\obin_k(\omega)
\nonumber \\ & \quad
  - 2\rabiP_{k} \left[ \oBB_{k}(\omega) - (2\rabiP_{k}/c|k|) \oXX_{k}(\omega) \right].
\end{align}
\end{subequations}
In both forms, the four equations are reduced to
\begin{equation}
\left[ c^2k^2 - \diep(\omega)\omega^2 \right] \oAA_k(\omega)
= \frac{4\omega\rabia \sqrt{\wa{}^3c|k|}}{\wa{}^2-\ii\gamma\wa - \omega^2}\sqrt{\gamma}\obin_k(\omega),
\end{equation}
where the dielectric function (dispersion relation) of the medium
is obtained as
\begin{equation} \label{eq:diep} 
\frac{c^2k^2}{\omega^2}
= \diep(\omega)
= 1 + \frac{4\rabia^2\wa{}^2}{\wa{}^2-\ii\gamma\wa-\omega^2}.
\end{equation}
From this dielectric function $\diep(\omega)$,
the spatial dependence of $\die(z,\omega)$ in Eq.~\eqref{eq:die(z,w)},
and by the MBCs,
we can calculate the absorption and reflection spectra of the cavity
embedding the dispersive and absorptive medium.

\subsection{Hamiltonian inside the cavity} \label{sec:oHz}
Whereas the spatially infinite system is considered
for deriving the dielectric function $\diep(\omega)$
of the medium in the previous sub-section,
here we show the Hamiltonian inside the cavity.
Since the bosonic excitations have an infinite mass,
we have a freedom of choosing the basis for describing the eigen-modes of them.
We expand the excitations by the same wavefunctions
as the photon modes in perfect cavity,
i.e., characterized by the confinement wavenumber $k_j$
in Eq.~\eqref{eq:k_j},
while the transition frequency is $\wa$ for all the excitation modes.
In the similar manner as the Hamiltonian $\oHcav$
for the empty cavity in Eq.~\eqref{eq:oHcav},
the Hamiltonian inside the filled cavity is described as
\begin{subequations} \label{eq:oHz_vr} 
\begin{align}
\frac{\oHCz}{\hbar}
& = \sum_{j} \left[
    ck_j \oad_{j}\oa_{j}
  + \wa \obd_j \ob_j
  + \rabiC_{j} \oAA_j \oYY_j
  + \frac{\rabiC_j{}^2}{\wa} \oAA_j \oAA_j
    \right], \\
\frac{\oHPz}{\hbar}
& = \sum_{j} \left[
    ck_j \oad_{j}\oa_{j}
  + \wa \obd_j \ob_j
  - \rabiP_{j} \oBB_j \oXX_j
  + \frac{\rabiP_{j}{}^2}{ck_j} \oXX_{j} \oXX_{j}
  \right].
\end{align}
\end{subequations}
The former and latter Hamiltonians are in the velocity and length forms,
respectively.
$\ob_j$ is the annihilation operator of an excitation with $k_j$.
The capital operators are defined as similar as in Eqs.~\eqref{eq:oABXY}.
The light-matter interaction strengths $\rabiC$ and $\rabiP$
are expressed by the non-dimensional strength $\rabia$ defined in Eq.~\eqref{eq:rabi} as
\begin{equation}
\rabiC_j
= \rabia \wa \sqrt{\frac{\wa}{ck_j}}, \quad
\rabiP_j
= \rabia ck_j \sqrt{\frac{\wa}{ck_j}}.
\end{equation}
In the similar manner as in the previous sub-section,
the SEC Hamiltonian for the excitation damping is supposed as
\begin{align}
\oHdamp
& = \sum_j\int_0^{\infty}\dd\omega\left\{
    \hbar\omega\ofd_j(\omega)\of_j(\omega)
\right. \nonumber \\ & \quad \left.
  + \ii\hbar\sqrt{\frac{\gamma}{2\pi}}
    \left[ \ofd_j(\omega) - \of_j(\omega) \right] \oX_j
  \right\}.
\end{align}

In the numerical calculation of the absorption spectra
by the Lindblad-type quantum Langevin equation,
we need the lowering and raising components
of the operators $\oAA_j, \oXX_j, \ldots$
Since the system of interest in this paper has no anharmonicity,
we can easily diagonalize the Hamiltonian $\oHz^{v/r}$ in Eqs.~\eqref{eq:oHz_vr}
by the Bogoliubov transformation \cite{Ciuti2005PRB,hopfield58}.
Since the confinement wavenumber $k_j$ is a good quantum number,
the system can be diagonalized for each $k_j$
by the polariton operator expressed as
\begin{equation}
\op_{j,\zeta} = w_{j,\zeta}\oa_j + x_{j,\zeta}\ob_j + y_{j,\zeta}\oad_j + z_{j,\zeta}\obd_j.
\end{equation}
For each $k_j$, there are lower mode ($\zeta = L$) and upper one ($\zeta = U$).
The coefficients $\{w_{j,\zeta}, x_{j,\zeta}, y_{j,\zeta}, z_{j,\zeta}\}$
and the eigen-frequencies $\{\omega_{j,\zeta}\}$ are determined
for satisfying
\begin{equation}
\left[ \op_{j,\zeta}, \oH_0^{v/r} \right] = \hbar\omega_{j,\zeta}\op_{j,\zeta}
\end{equation}
and the normalization condition
\begin{equation}
\left[ \op_{j,\zeta}, \opd_{j',\zeta'} \right] = \delta_{j,j'}\delta_{\zeta,\zeta'}.
\end{equation}
For the velocity form,
the detailed analytical expressions are shown
in Ref.~\cite{hopfield58,Bamba2013MBC}.
For the velocity and length forms,
we get the same eigen-frequencies $\{\omega_{j,\zeta}\}$,
while the coefficients $\{w_{j,\zeta}, x_{j,\zeta}, y_{j,\zeta}, z_{j,\zeta}\}$
are different.
In the calculations in this paper,
we checked numerically that the two forms certainly give the same results.
Since the annihilation operators are expressed as
\begin{subequations}
\begin{align}
\oa_j & = \sum_{\zeta = L,U}\left( w_{j,\zeta}^*\op_{j,\zeta} - y_{j,\zeta}\opd_{j,\zeta} \right), \\
\ob_j & = \sum_{\zeta = L,U}\left( x_{j,\zeta}^*\op_{j,\zeta} - z_{j,\zeta}\opd_{j,\zeta} \right),
\end{align}
\end{subequations}
the lowering components of the Hermitian operators are represented such as
\begin{subequations} \label{eq:oAAp_p} 
\begin{align}
\oAAp_j & = \sum_{\zeta = L,U} (w_{j,\zeta}^*-y_{j,\zeta}^*)\op_{j,\zeta}, \\
\oXXp_j & = \sum_{\zeta = L,U} (x_{j,\zeta}^*-z_{j,\zeta}^*)\op_{j,\zeta}.
\end{align}
\end{subequations}

\section{Calculations by three approaches} \label{sec:calculation}
In order to evaluate the validity of the Lindblad and non-Lindblad master equations,
we compare the absorption spectra calculated by them
(exactly speaking, by corresponding quantum Langevin equations)
and that by the MBCs along the classical electrodynamics.
The comparison and discussion will be performed in Sec.~\ref{sec:compare}.
In this section,
we show the calculation methods of the three approaches:
by the MBCs in Sec.~\ref{sec:byMBCs},
non-Lindblad-type equation in Sec.~\ref{sec:by_non-Lindblad},
and Lindblad-type one in Sec.~\ref{sec:by_Lindblad}.

\subsection{Absorption by Maxwell boundary conditions} \label{sec:byMBCs}
From the spatially dependent dielectric function $\die(z,\omega)$
in Eq.~\eqref{eq:die(z,w)} and $\diep(\omega)$ of the medium
in Eq.~\eqref{eq:diep},
we can calculate the reflection and absorption spectra
of the cavity system.
The electric field in the whole system is expressed as
\begin{equation}
E(z,\omega)
= \begin{cases}
E_{0}(\omega) \ee^{\ii(\omega/c)z} + E_{r}(\omega) \ee^{-\ii(\omega/c)z} & z < 0 \\
E_1(\omega) \sin[\kp(\omega)(\cavlen-z)] & 0 < z < \cavlen
\end{cases}
\end{equation}
Here, $E_0(\omega)$, $E_r(\omega)$, and $E_1(\omega)$
are the electric field of the incident wave,
of reflected wave, and inside the cavity, respectively.
This expression satisfies the MBC at $z = \cavlen$ as
$E(\cavlen,\omega) = 0$.
The complex wavenumber $\kp(\omega)$ inside the cavity
is defined with the refractive index $\np(\omega) = \sqrt{\diep(\omega)}$ as
\begin{equation}
\kp(\omega) = \np(\omega) \omega / c.
\end{equation}
From the MBCs at $z = 0$ in Eqs.~\eqref{eq:MBC}, we get
\begin{subequations}
\begin{align}
E_0(\omega) + E_r(\omega) & = E_1(\omega) \sin[\kp(\omega)\cavlen], \\
E_0(\omega) - E_r(\omega)
& = \ii \{ \np(\omega) \cos[\kp(\omega)\cavlen]
\nonumber \\ & \quad
- \varLambda(\omega)\sin[\kp(\omega)\cavlen] \} E_1(\omega).
\end{align}
\end{subequations}
Then, the reflection coefficient $r(\omega) = E_r(\omega)/E_0(\omega)$
is obtained as
\begin{equation} \label{eq:r(w)} 
r(\omega)
= \frac{[1+\ii\varLambda(\omega)]\sin[\kp(\omega)\cavlen] - \ii\np(\omega)\cos[\kp(\omega)\cavlen]}
         {[1-\ii\varLambda(\omega)]\sin[\kp(\omega)\cavlen] + \ii\np(\omega)\cos[\kp(\omega)\cavlen]}.
\end{equation}
The reflectance and absorption are calculated as
$R(\omega) = |r(\omega)|^2$ and $1-R(\omega)$,
respectively.

\begin{figure}[htbp]
\includegraphics[width=.9\linewidth]{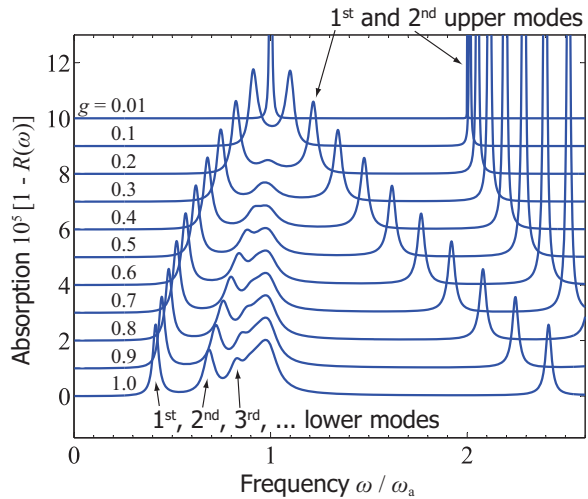}
\caption{
Absorption $1-R(\omega)$ calculated by the MBCs is plotted
as a function of $\omega/\wa$.
The light-matter interaction strength is changed
as $\rabi = 0.01$, 0.1, 0.2, \ldots, 1.0.
Parameters: $\cavlen = \pi c/\wa$, $\gamma = 0.1\wa$,
and $\varLambda_0 = 10^3$.}
\label{fig:2}
\end{figure}
In Fig.~\ref{fig:2}, we plot the absorption spectra $1-R(\omega)$
calculated by the MBCs with changing the light-matter interaction strength
$\rabi = 0.01$, 0.1, 0.2, \ldots, 1.0.
The cavity length is supposed as a half of the light wavelength at $\wa$
in vacuum: $\cavlen = \pi c /\wa$.
Then, the resonance frequency of the lowest cavity mode ($j = 1$)
is $c k_1 = \wa$.
We supposed that the damping rate is $\gamma = 0.1\wa$
and the cavity loss is $\varLambda_0 = 10^3$,
which corresponds to $\kappa_j = j\times6.366\times10^{-7}\wa$
for the $j$-th cavity mode,
and the quality factor is $Q = c k_j/\kappa_j = 1.571\times10^6$
for the empty cavity.
We basically suppose such a good cavity,
because the SEC Hamiltonian in Eq.~\eqref{eq:oHSECcav}
is valid only for good cavities.
For bad cavities, while the absorption spectra
can be calculated by the MBCs for given $\eta(\omega)$ or $\varLambda(\omega)$,
the master and quantum Langevin equations
cannot well reproduce them due to the invalidity of the SEC Hamiltonian.

Since the damping rate is $\gamma = 0.1\wa$,
$\rabi = 0.01$ corresponds to the weak interaction regime,
and the absorption peaks are found at $\omega = c k_j$
on the uppermost spectrum in Fig.~\ref{fig:2}.
For larger $\rabi$, we can find the peak splitting
of the lowest (first) cavity mode and the excitations
as the lower and upper polariton modes.
In the ultra-strong interaction regime $\rabi \sim 1$,
the center of these two peaks is shifted to the higher frequency side,
and the lower polariton frequency never becomes a negative or imaginary value.
Further, the peaks of the second (third) lower polariton mode with $k_2$ ($k_3$)
gradually appears, because the photonic component
of these lower polariton modes are increased
(excited more efficiently) by the increase in $\rabi$.
For the present parameters,
the lower polariton modes with $j > 3$ appears
as a broad and asymmetric peak around $\omega \sim \wa$.

\begin{figure}[htbp]
\includegraphics[width=.9\linewidth]{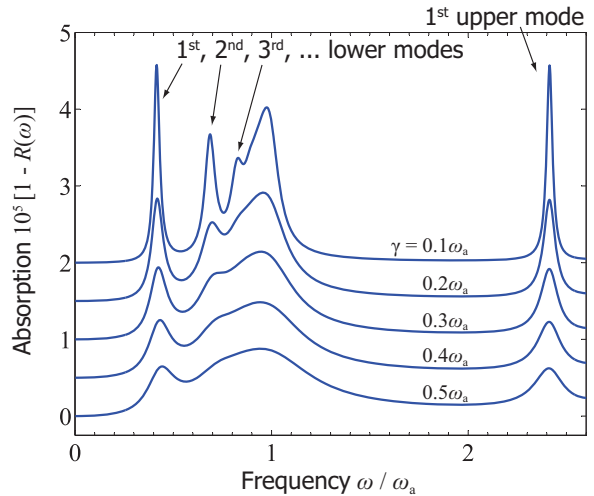}
\caption{Absorption $1-R(\omega)$ calculated by the MBCs is plotted
as a function of $\omega/\wa$.
The damping rate $\gamma$ is changed from $0.1\wa$ to $0.5\wa$.
Parameters: $\cavlen = \pi c/\wa$, $\rabi = 1$,
and $\varLambda_0 = 10^3$.}
\label{fig:3}
\end{figure}
In Fig.~\ref{fig:3}, we fixed the light-matter interaction strength as $\rabi = 1$,
but the damping rate $\gamma$ is changed from $0.1\wa$ to $0.5\wa$.
Increasing $\gamma$, the absorption peaks become broadened.
For $\gamma = 0.5\wa$, we cannot clearly find the second and the third
lower polariton modes.

In the following discussion, we will use the undermost absorption spectrum
in Fig.~\ref{fig:3}
as a standard reference for evaluating the validity
of the Lindblad and non-Lindblad master equations.
This is because the discrepancy between them
are highlighted for large $\rabi$ and $\gamma$ as will be found
in Sec.~\ref{sec:compare}.

\subsection{Absorption by non-Lindblad-type equations} \label{sec:by_non-Lindblad}
For the total Hamiltonian
$\oH = \oHz + \oHSECcav + \oHenvcav + \oHdamp$,
the quantum Langevin equation corresponding to the non-Lindblad master equation
is obtained as
\begin{widetext}
\begin{align} \label{eq:Langevin_non-Lindblad_cav_damp} 
\ddt{}\oS
& = \frac{1}{\ii\hbar}\left[ \oS, \oH_0 \right]
- \sum_j \int_0^{\infty}\dd\omega\ \left\{
  \left[ \oS, \oXX_j \right] \ee^{-\ii\omega t}
  \left[\frac{\gamma}{2}\oXX_j(\omega) + \sqrt{\gamma}\obin_j(\omega)\right]
  + \Hc
  \right\}
\nonumber \\ & \quad
- \int_0^{\infty}\dd\omega\ \left\{
  \sum_j \left[ \oS, \oAA_j \right] \ee^{-\ii\omega t}\sqrt{\kappa_j}
  \left[ \oain(\omega) + \sum_{j'}\frac{\sqrt{\kappa_{j'}}}{2}\oAA_{j'}(\omega) \right]
  + \Hc
  \right\}.
\end{align}
\end{widetext}
The equations of motion of the positive-frequency components of
four Hermitian operators are obtained in the velocity form as ($\omega > 0$)
\begin{subequations} \label{eq:ABXY_nonLindblad} 
\begin{align}
-\ii\omega\oAA_{j}(\omega) & = - ck_j \oBB_{j}(\omega), \\
-\ii\omega\oBB_{j}(\omega)
& = (ck_j+4\rabiC_{j}{}^2/\wa) \oAA_{j}(\omega)
    + 2\rabiC_{j}\oYY_{j}(\omega)
\nonumber \\ & \quad
  - \sum_{j'}\ii\sqrt{\kappa_j\kappa_{j'}}\oAA_{j'}(\omega)
  - \ii2\sqrt{\kappa_j}\oain(\omega), \\
-\ii\omega\oXX_{j}(\omega) & = -\wa\oYY_{j}(\omega) - 2\rabiC_{j}\oAA_{j}(\omega), \\
-\ii\omega\oYY_{j}(\omega) & = (\wa-\ii\gamma) \oXX_{j}(\omega) - \ii2\sqrt{\gamma}\obin_j(\omega).
\end{align}
\end{subequations}
The wavenumber $k_j$ is no longer a good quantum number
due to the coupling between inside and outside the cavity.
This equation set can be solved numerically,
and we can get the expression of $\oAA_j(\omega)$ such as
\begin{equation}
\oAA_j(\omega) = \alpha_j(\omega)\oain(\omega) + \sum_{j'} \beta_{j,j'}(\omega)\obin_{j'}(\omega),
\end{equation}
where the coefficients $\alpha_j(\omega)$ and $\beta_{j,j'}(\omega)$
are determined numerically.
Since the input-output relation is written as in Eq.~\eqref{eq:inout_non-Lindblad},
the output operator is represented as
\begin{equation}
\oaout(\omega) = r(\omega) \oain(\omega)
+ \sum_{j,j'} \sqrt{\kappa_j}\beta_{j,j'}(\omega)\obin_{j'}(\omega),
\end{equation}
where the reflection coefficient is expressed as
\begin{equation}
r(\omega) = 1 + \sum_j \sqrt{\kappa_j} \alpha_j(\omega).
\end{equation}
The reflection is calculated as $R_{\text{non-Lindblad}}(\omega) = |r(\omega)|^2$,
and the absorption is $1 - R_{\text{non-Lindblad}}(\omega)$.
We checked numerically that the same results are obtained even in the length form.

\subsection{Absorption by Lindblad-type equations} \label{sec:by_Lindblad}
The total Hamiltonian
is written as $\oH = \oHz + \oHSECcavRWA + \oHenvcav + \oHdamp$.
In the quantum Langevin equation,
the cavity loss is introduced corresponding to the Lindblad master equation,
while the damping is treated in the non-Lindblad form.
The quantum Langevin equation is expressed in this case
(under the RWA to the cavity loss) as
\begin{widetext}
\begin{align} \label{eq:Langevin_Lindblad_cav_damp} 
\ddt{}\oS
& = \frac{1}{\ii\hbar}\left[ \oS, \oH_0 \right]
- \sum_j \int_0^{\infty}\dd\omega\ \left\{
  \left[ \oS, \oXX_j \right] \ee^{-\ii\omega t}
  \left[\frac{\gamma}{2}\oXX_j(\omega) + \sqrt{\gamma}\obin_j(\omega)\right]
  + \Hc
  \right\}
\nonumber \\ & \quad
- \int_0^{\infty}\dd\omega\ \left\{
  \sum_j \left[ \oS, \oAAm_j \right] \ee^{-\ii\omega t}\sqrt{\kappa_j}
  \left[ \oain(\omega) + \sum_{j'}\frac{\sqrt{\kappa_{j'}}}{2}\oAAp_{j'}(\omega) \right]
  + \Hc
  \right\}.
\end{align}
\end{widetext}
The commutator $[\oS, \oAAm_j]$ is calculated
by rewriting $\oS$ with the polariton operators
$\{\op_{j,\zeta}, \opd_{j,\zeta}\}$
and by rewriting $\oAAm_j$ with $\{\opd_{j,\zeta}\}$
as in Eqs.~\eqref{eq:oAAp_p}.
We calculated the absorption spectra in the similar manner
as in the previous sub-section,
while the detailed equations are shown in App.~\ref{app:eq_set}.

In this calculation,
we replaced the Fourier transform of the lowering component
$\oAAp_{j'}(\omega)$ by the positive-frequency component
$\oAA_{j'}(\omega)$ for $\omega > 0$.
We checked numerically the validity of this replacement
in the case that both the cavity loss and damping
is treated in the Lindblad form.
This is because we can easily calculate the absorption spectra
by replacing all the positive-frequency components of Hermitian operators
with the Fourier transform of the lowering components,
e.g., $\oAA_{j}(\omega)$ is replaced by $\oAAp_{j}(\omega)$,
and polariton annihilation operators $\{\op_{j,\zeta}\}$
and creation ones $\{\opd_{j,\zeta}\}$ are not mixed 
in the equations of motion.
We checked numerically that the absorption spectra by these approaches
are approximately equivalent.
The detail of this discussion is shown in App.~\ref{app:eq_set}.

\section{Comparison of three approaches} \label{sec:compare}
\begin{figure*}[tbp]
\tabcolsep = .5em
\begin{tabular}{rl}
\begin{minipage}[t]{.47\textwidth}
\includegraphics[scale=.4]{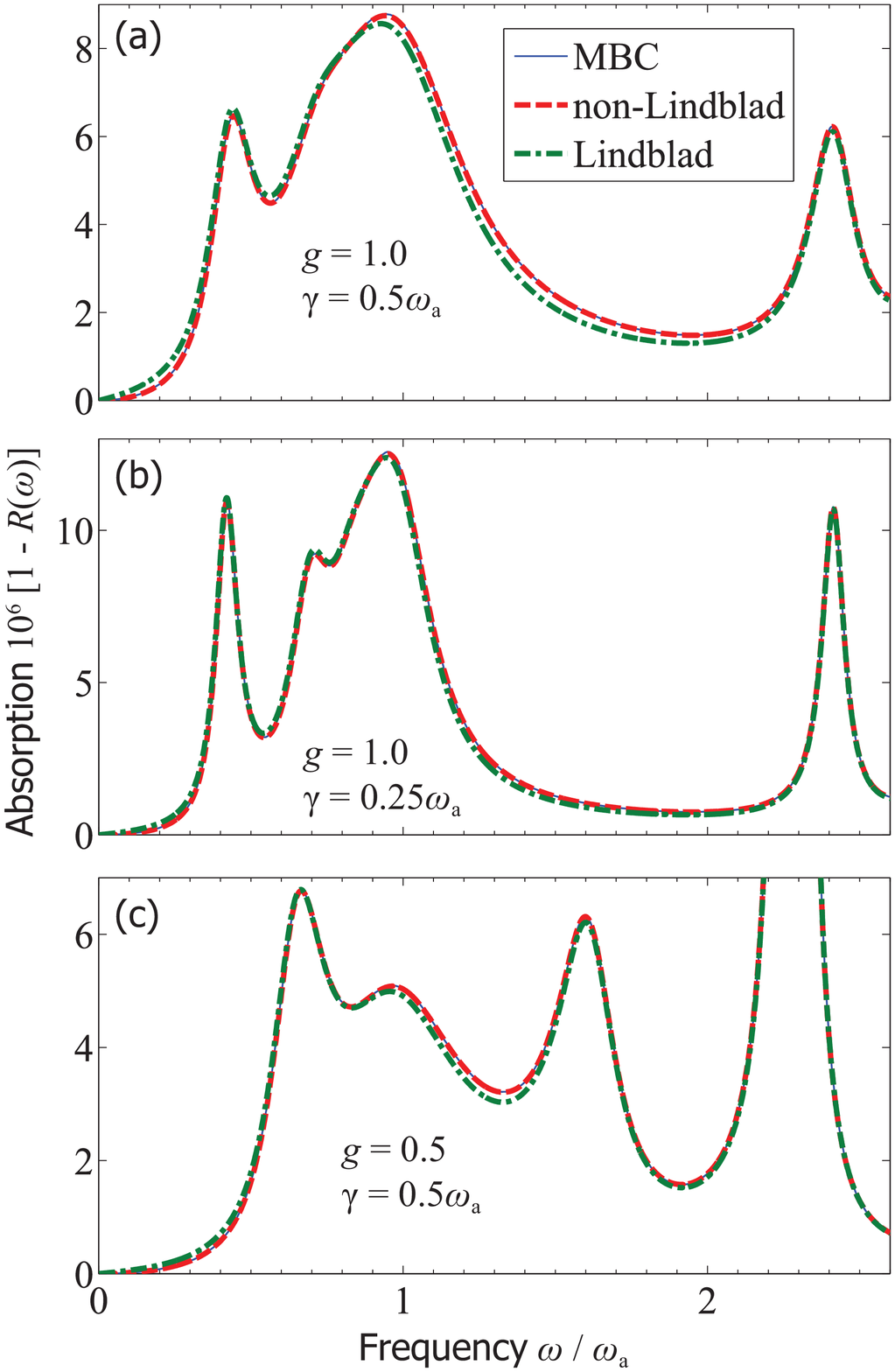}
\caption{Absorption spectra by three approaches:
by the MBCs (blue solid line),
by non-Lindblad-type equation (red dashed line),
and by Lindblad-type equation (green dash-dotted line).
We considered (a) $\rabi = 1.0$ and $\gamma = 0.5\wa$,
(b) $\rabi = 1.0$ and $\gamma = 0.25\wa$, and
(c) $\rabi = 0.5$ and $\gamma = 0.5\wa$.
We basically get good agreement between the spectra
by the MBCs and by the non-Lindblad-type equation,
while those by the Lindblad-type equation show a discrepancy.
The discrepancy is basically reduced by the decrease in $\rabi$ and $\gamma$.
Parameters: $\cavlen = \pi c/\wa$ and $\varLambda_0 = 10^3$.
$k_j$ up to $j = 2000$ are considered.}
\label{fig:4}
\end{minipage} &
\begin{minipage}[t]{.47\textwidth}
\includegraphics[scale=.4]{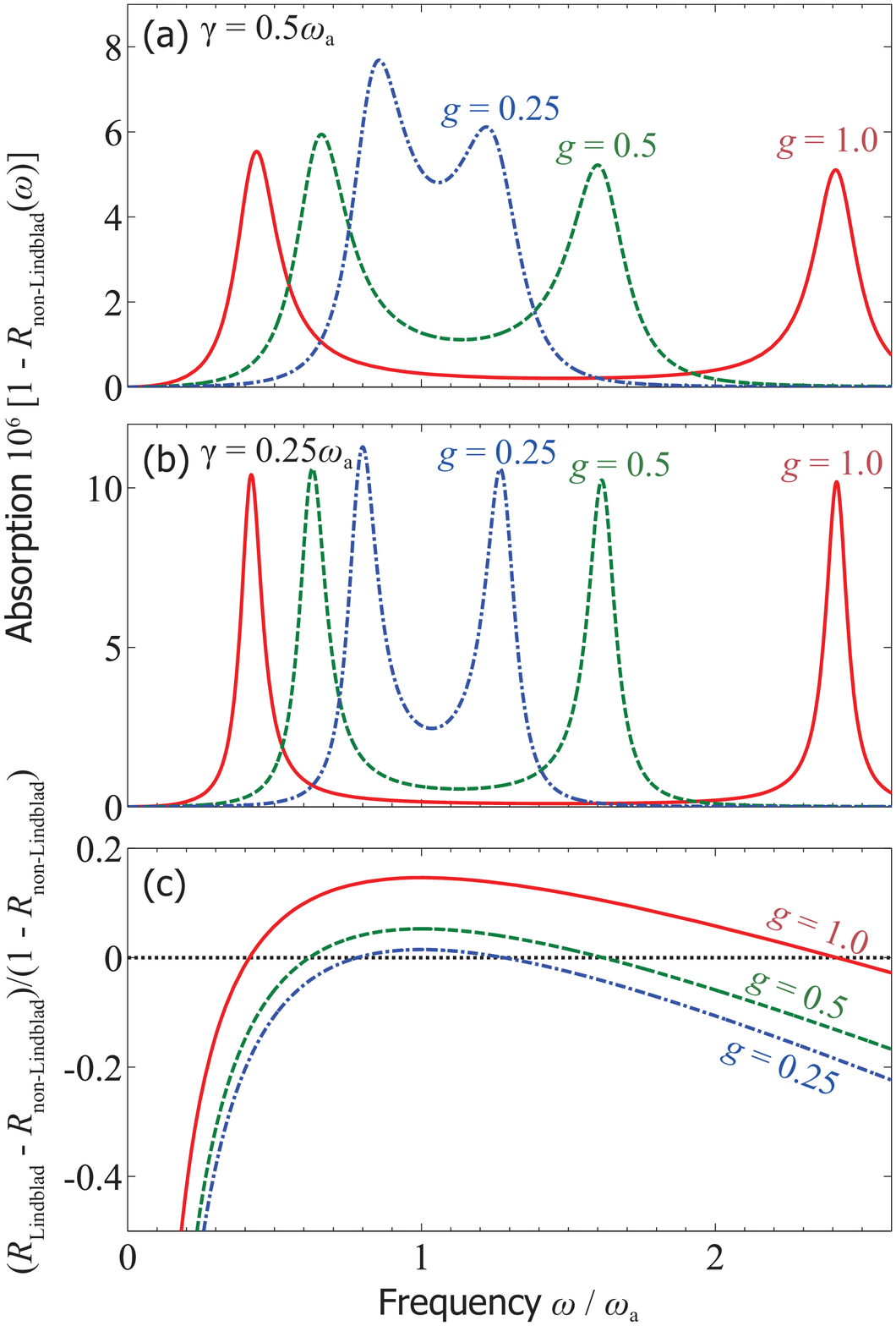}
\caption{(a,b) Absorption spectra calculated by the non-Lindblad-type equation
by considering only the lowest wavenumber $k_1$
are plotted  for $g = 1.0$ (red solid line),
$g = 0.5$ (green dashed line), and $g = 0.25$ (blue dash-dotted line).
We supposed (a) $\gamma = 0.5\wa$ and (b) $\gamma = 0.25\wa$.
(c) The differences of absorption spectra by the Lindblad-
and non-Lindblad-type equations normalized to that of the non-Lindblad-type one
are plotted as functions of $\omega/\wa$.
These curves depend only on $\cavlen$ and $\rabi$.
Parameters: $\cavlen = \pi c/\wa$ and $\varLambda_0 = 10^3$.}
\label{fig:5}
\end{minipage}
\end{tabular}
\end{figure*}
In Fig.~\ref{fig:4}(a), we plot the absorption spectra
by the three approaches: the MBCs (solid line),
non-Lindblad-type quantum Langevin equation (dashed line),
and Lindblad-type one (dash-dotted line).
The parameters are the same as the undermost spectrum in Fig.~\ref{fig:3}:
$\cavlen = \pi c/\wa$, $\varLambda_0 = 10^3$, $\rabi = 1$,
and $\gamma = 0.5\wa$.
In the calculations by the quantum Langevin equations,
we considered the wavenumbers $\{k_j\}$ up to $j = 2000$.

We can clearly find that the spectra by the MBCs and non-Lindblad-type equation
agree well with each other.
A small discrepancy is found at the top of the broad peak around $\omega = \wa$.
This broad peak appears as the sum of the peaks of lower polariton modes
for $2 < j < \infty$.
The small discrepancy basically comes from our neglect of higher modes for $j > 2000$
in the calculation based on the non-Lindblad-type quantum Langevin equation.

On the other hand, the spectrum by the Lindblad-type equation
clearly shows a discrepancy with that by the other two approaches.
Basically, a larger absorption is obtained for $\omega < \wa$,
and smaller is obtained for $\omega > \wa$ in the frequency range of the figure.
It is hard to explain this discrepancy in terms of physics,
but, mathematically, it comes from the difference
between the commutator $[ \oS, \oAA_j ]$
at the dissipation and noise terms concerning the cavity loss
in the non-Lindblad-type equation
\eqref{eq:Langevin_non-Lindblad_cav_damp}
and $[ \oS, \oAAp_j ]$ in the Lindblad-type equation
\eqref{eq:Langevin_Lindblad_cav_damp}.
This difference originates from the RWA to the SEC.

In order to catch the tendency of this discrepancy,
in Fig.~\ref{fig:4}(b),
we decreased the damping rate as $\gamma = 0.25\wa$
while keeping the other parameters $\rabi$, $\varLambda_0$, and $\cavlen$.
All the absorption peaks become narrower than those in Fig.~\ref{fig:4}(a),
and we can now find the peak originating from the second lower polariton mode.
The discrepancy between the spectra by the Lindblad-type equation
and the other two is reduced than Fig.~\ref{fig:4}(a),
although we did not change the cavity-loss parameter $\varLambda_0$
and the RWA was performed to the SEC concerning the cavity loss.

On the other hand, in Fig.~\ref{fig:4}(c),
we instead decreased the light-matter interaction strength as $\rabi = 0.5$
than in Fig.~\ref{fig:4}(a)
while keeping $\gamma$, $\varLambda_0$, and $\cavlen$.
The discrepancy becomes relatively smaller than Fig.~\ref{fig:4}(a)
especially around the top of the peaks
and the tail of the lowest peak,
while we can still find the discrepancy around $\omega \sim \wa$.

In order to understand these tendencies of the discrepancy,
we tentatively simplify the calculation,
i.e., let us consider only the lowest wavenumber $k_1$ for photons and excitations
in the Lindblad and non-Lindblad quantum Langevin equations.
Since we cannot eliminate the higher $k_j$ modes
in the calculation by the MBCs,
we compare the absorption spectra by the two quantum Langevin equations.

In Fig.~\ref{fig:5}(a),
we plot the absorption spectra obtained by the non-Lindblad-type
quantum Langevin equation for $\rabi = 1.0$, 0.5, and 0.25
with keeping $\gamma = 0.5\wa$, $\varLambda_0 = 10^3$, and $\cavlen = \pi c/\wa$.
The broad peak around $\omega \sim \wa$ disappears,
and we simply get the peaks of the lower and upper polariton modes with $k_1$.
In Fig.~\ref{fig:5}(b), we plot the absorption spectra
in the case of $\gamma = 0.25\wa$.
We can find narrower peaks than in Fig.~\ref{fig:5}(a).
Concerning the discrepancy between the Lindblad- and non-Lindblad-type equations,
we plot the normalized absorption difference
$(R_{\text{Lindblad}}-R_{\text{non-Lindblad}})/(1-R_{\text{non-Lindblad}})$
for $\rabi = 1.0$, 0.5, and 0.25 in Fig.~\ref{fig:5}(c).
Surprisingly, these curves are independent of $\gamma$ and $\varLambda_0$,
but they depend on only $\rabi$ and $\cavlen$.
We get positive values between the lower and upper polariton frequencies,
and negative values are obtained out of this frequency region.
Especially, the normalized difference diverges for $\omega \rightarrow 0$.
This is because the original lowering components such as $\oAAp_j$
and the raising ones $\oAAm_j$ are mixed through the SEC
for $\omega \ll \wa$, because the frequency difference $2\omega$
of them becomes negligible than the light-matter interaction strength $\rabi\wa$.
However, since the absorption peaks drop well to zero for $\omega \rightarrow 0$,
we do not focus on this divergence.

Since the normalized difference is zero at the lower and upper polariton frequencies,
we got the smaller discrepancy in Fig.~\ref{fig:4}(b)
by the decrease in the broadening $\gamma$.
The discrepancy is basically highlighted for large $\gamma$
especially when $\gamma$ is comparable to or larger than the mode splitting
$\sim 2\rabi\wa$ in this case.
On the other hand, in Fig.~\ref{fig:5}(c), the top of the curves at $\omega = \wa$
are 0.1464, 0.0528, and 0.0149 for $\rabi = 1.0$, 0.5, and 0.25, respectively.
Thanks to this reduction of the discrepancy with the decrease in $\rabi$,
we get the smaller discrepancy at the top and the lower tail
of the lower polariton peak in Fig.~\ref{fig:4}(c)
than in Fig.~\ref{fig:4}(a).
For understanding the remaining discrepancy around $\omega = \wa$ in Fig.~\ref{fig:4}(c),
we need to consider also the higher $k_j$ modes for $j > 1$.
We calculated the absorption spectra
by the Lindblad- and non-Lindblad-type quantum Langevin equations
with eliminating the mixing of different $k_j$,
i.e, the summation over $j'$ is performed only for $j' = j$
in Eqs.~\eqref{eq:Langevin_non-Lindblad_cav_damp}
and \eqref{eq:Langevin_Lindblad_cav_damp}.
Although we do not show the numerical results in figures,
we got almost the same absorption spectra
as in Fig.~\ref{fig:4}, which was calculated with the $k_j$ mixing.
This minor contribution of the $k_j$ mixing is natural
because we supposed the good cavity with $\varLambda_0 \gg 1$.
Then, we can say that the discrepancy in Fig.~\ref{fig:4}
basically comes from the summation of the differences
seen in Fig.~\ref{fig:5}(c) for all the $k_j$ modes.

Concerning this discrepancy in Fig.~\ref{fig:4}(c),
whereas the $k_1$ modes is a part of its origin,
the higher $k_j$ modes also contribute to it.
From these facts, we can say that
the discrepancy between the Lindblad-type equation
and the other two methods clearly appear
in the ultra-strong light-matter interaction regime
[as seen in Fig.~\ref{fig:5}(c)]
with a concentration of modes
(overlap of absorption peaks),
which is increased by enlarging $\gamma$, i.e., wider broadening.


Anyway, whereas we considered only the good cavity case ($\varLambda_0 \gg 1$),
we basically found a good agreement between the absorption spectra
by the MBCs and the non-Lindblad-type quantum Langevin equation.
The absorption spectrum by the Lindblad-type equation
shows a discrepancy from them
in the ultra-strong light-matter interaction regime
with a large broadening $\gamma$ (overlap).

If the cavity quality is not so good as $\varLambda_0 \lesssim 1$,
we cannot reproduce the absorption spectrum of the MBCs
even by the non-Lindblad-type quantum Langevin equation.
This is because our SEC Hamiltonian for the cavity loss
is valid only for the good cavities.
On the other hand, for much low damping $\gamma \ll \wa$,
the peak widths are found to be almost the same for the three approaches.
However, since we did not consider the Lamb shift due to the SEC,
the Lindblad and non-Lindblad master equations
give almost the same absorption spectra,
while the peak positions are found to be shifted in the spectra by the MBCs.

From these facts, we conclude that we should not apply the RWA to the SEC
of the cavity loss in the derivation of the master and quantum Langevin equations.
Although the master equation of the non-Lindblad form is derived,
it shows a better agreement with calculations by the MBCs in the classical electrodynamics,
comparing to the Lindblad master equation derived under the RWA to the SEC.
Although we did not discuss in detail the other SECs, such as the damping,
the same conclusion is probably obtained
by considering explicitly the mechanism of the SECs,
because the RWA is an approximation basically for the mathematical requirement,
i.e., the positivity of density operator in the Lindblad master equation.

\section{Positivity check} \label{sec:positivity}
Next, we check the violation of the positivity in the non-Lindblad master equation.
The non-Lindblad-type quantum Langevin equation
in Eq.~\eqref{eq:Langevin_non-Lindblad_cav_damp}
used in the previous sections
corresponds to the following non-Lindblad master equation:
\begin{align} \label{eq:master_kappa_gamma} 
\ddt{}\orho
& = \frac{1}{\ii\hbar}\left[ \oHz, \orho \right]
  + \sum_{j}
    \frac{\gamma}{2}
      \left(
        \left[ \oXX_j, \orho \oXXm_{j} \right]
      + \left[ \oXXp_{j}\orho, \oXX_j \right]
      \right)
\nonumber \\ & \quad
  + \sum_{j,j'}
    \frac{\sqrt{\kappa_j\kappa_{j'}}}{2}
      \left(
        \left[ \oAA_j, \orho \oAAm_{j'} \right]
      + \left[ \oAAp_{j'}\orho, \oAA_j \right]
      \right).
\end{align}
For the ground state $\ketg$ of the Hamiltonian $\oHz$
of the system of interest,
it is clear that $\rho = \ketg\brag$ is a steady state solution
of Eq.~\eqref{eq:master_kappa_gamma},
because $\oXXp_j\ketg = \oAAp_j\ketg = 0$
and $\brag\oXXm_j = \brag\oAAm_j = 0$.
Then, the positivity is never violated
in the steady state in this dissipative situation.

We therefore check the positivity in non-equilibrium or dynamical situations
by numerically solving the non-Lindblad master equation.
However, due to the computational difficulty,
we here consider only the lowest cavity and excitation modes ($j = 1$),
and the number of bosons in each mode is limited to 24,
which is large enough in the following calculations.
This computational cost is the reason why we used the quantum Langevin equations
in the previous sections,
where we could consider 2000 modes without limiting the number of bosons.
The non-Lindblad master equation used in this section is
\begin{align} \label{eq:master_kappa_gamma_n} 
\ddt{}\orho
& = \frac{1}{\ii\hbar}\left[ \oHz^{j=1}, \orho \right]
  + \frac{\gamma n}{2}
      \left(
        \left[ \oXX_1, \orho \oXXp_{1} \right]
      + \left[ \oXXm_{1}\orho, \oXX_1 \right]
      \right)
\nonumber \\ & \quad
  + \frac{\gamma(1+n)}{2}
      \left(
        \left[ \oXX_1, \orho \oXXm_{1} \right]
      + \left[ \oXXp_{1}\orho, \oXX_1 \right]
      \right)
\nonumber \\ & \quad
  + \frac{\kappa_1}{2}
      \left(
        \left[ \oAA_1, \orho \oAAm_{1} \right]
      + \left[ \oAAp_{1}\orho, \oAA_1 \right]
      \right).
\end{align}
The Hamiltonian $\oHz^{j=1}$
consists of only the lowest photonic and excitonic modes with $j = 1$.
We assume that the distribution of the excitonic environment is flat as
$n(\omega) = n$ in the frequency range of interest.

We first check the positivity in the non-equilibrium steady state
under the incoherent excitation by the excitonic environment with $n > 0$.
The density operator $\orhoss$ in the steady state is numerically calculated
by searching zero eigen-value of the coefficient matrix
for $\orho$ on the right-hand side in Eq.~\eqref{eq:master_kappa_gamma_n}.
As far as we checked numerically,
the minimum eigen-value of $\orhoss$ is basically zero
within the range of numerical error.

\begin{figure*}[tbp]
\tabcolsep = .5em
\begin{tabular}{rl}
\begin{minipage}[t]{.47\textwidth}
\includegraphics[scale=.4]{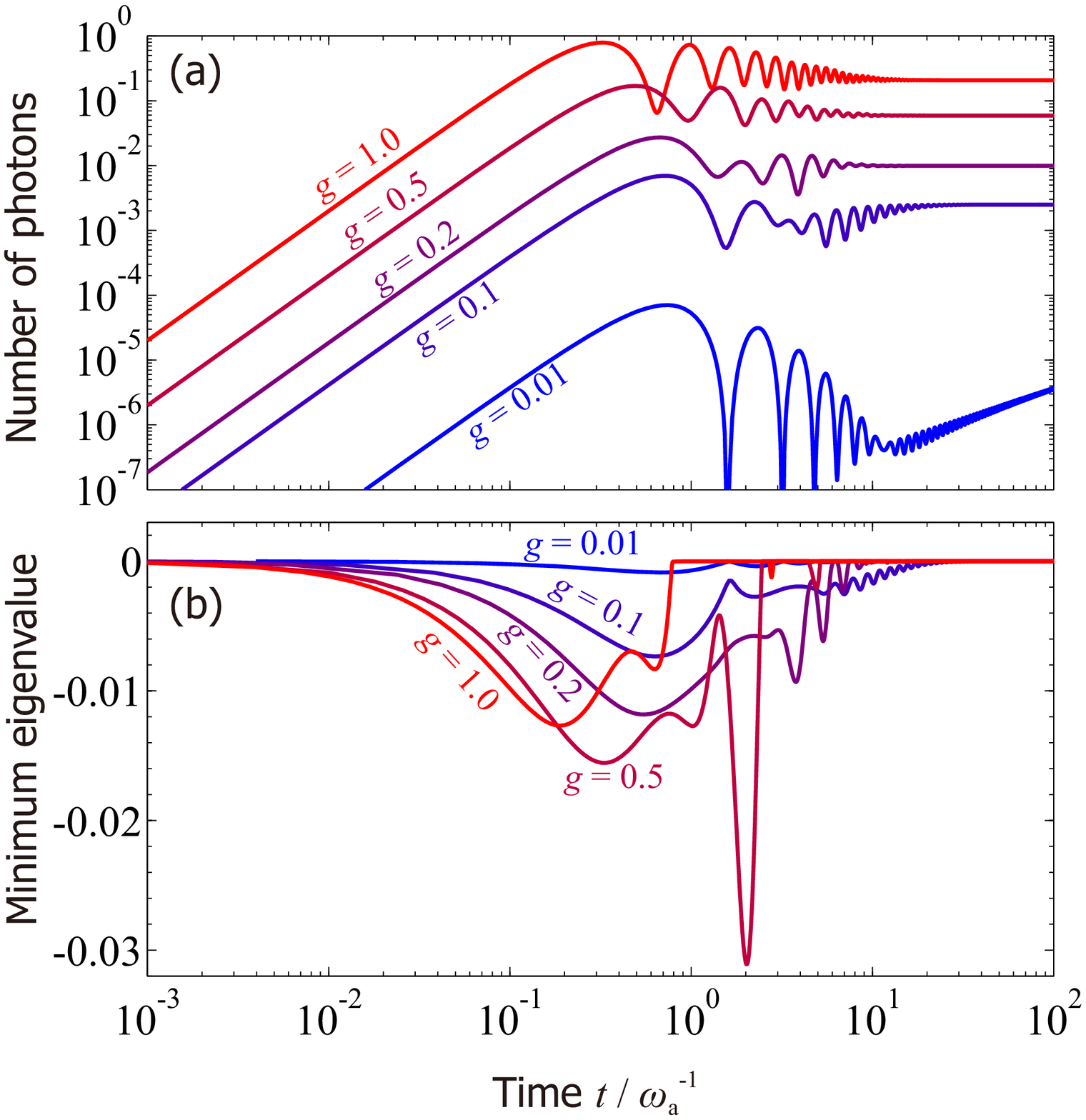}
\caption{
Temporal development of (a) number of photons inside the cavity
and (b) minimum eigen-value of density operator
calculated by the non-Lindblad master equation in Eq.~\eqref{eq:master_kappa_gamma_n}
for $n = 0$ (no incoherent excitation).
The initial density operator is $\orho(t=0) = \ket{0,0}\bra{0,0}$
with zero photon and zero excitation.
The results for $\rabi = 0.01$, 0.1, 0.2, 0.5, and 1.0 are plotted
with different color.
Negative eigen-values are clearly obtained as seen in panel (b),
while they become negligible when the system reaches the steady state
$\orhoss = \ketg\brag$.
Parameters: $ck_1 = \wa$, $\kappa_1 = 6.366\times10^{-7}\wa$,
$\gamma = 0.5\wa$, and $n = 0$.
}
\label{fig:6}
\end{minipage} &
\begin{minipage}[t]{.47\textwidth}
\includegraphics[scale=.4]{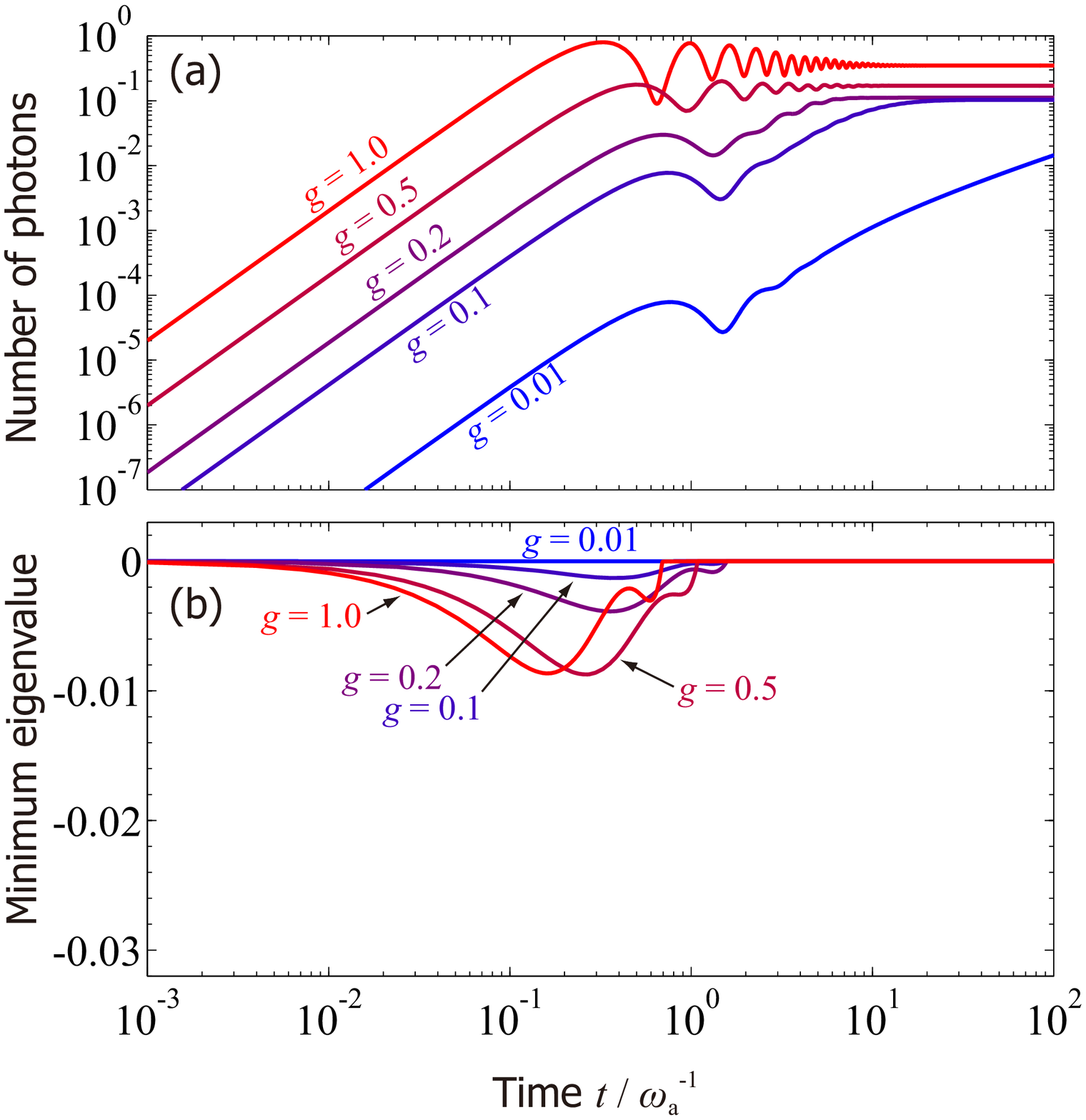}
\caption{
Temporal development of (a) number of photons inside the cavity
and (b) minimum eigen-value of density operator
calculated by the non-Lindblad master equation in Eq.~\eqref{eq:master_kappa_gamma_n}
for $n = 0.1$ (under incoherent excitation).
The initial density operator is $\orho(t=0) = \ket{0,0}\bra{0,0}$
with zero photon and zero excitation.
The results for $\rabi = 0.01$, 0.1, 0.2, 0.5, and 1.0 are plotted
with different color.
Comparing to Fig.~\ref{fig:6}(b),
the negative eigen-values are rapidly suppressed
due to the presence of of the incoherent excitation.
Parameters: $ck_1 = \wa$, $\kappa_1 = 6.366\times10^{-7}\wa$,
$\gamma = 0.5\wa$, and $n = 0.1$.
}
\label{fig:7}
\end{minipage}
\end{tabular}
\end{figure*}
Then, what we have to check is the positivity in dynamical situation.
Compared with starting from the ground state $\orho = \ketg\brag$
under the incoherent excitation ($n > 0$),
we can find a clear violation of the positivity
by starting from the state with zero photon and zero excitation
$\ketzz$,
which is not the ground state $\ketg$ in the ultra-strong interaction regime.

Figure \ref{fig:6}(a) shows the temporal development of the number of photons
starting from $\rho(t=0) = \ketzz\brazz$.
We supposed $c k_1 = \wa$, $\kappa_1 = 6.366\times10^{-7}\wa$,
$\gamma = 0.5\wa$, and no incoherent excitation $n = 0$.
The results for
$\rabi = 0.01$, 0.1, 0.2, 0.5, and 1.0 are plotted with different color.
Since $\ketzz$ is not the true ground state of $\oHz^{j=1}$,
the number of photons gradually increases
and oscillatory reaches $\brag\oad_1\oa_1\ketg$,
which means $\rho(t\to\infty) = \ketg\brag$.
In the case of relatively weak interaction strength $\rabi = 0.01$,
the density operator does not yet reach $\ketg\brag$ at $t = 100\wa^{-1}$.

Figure \ref{fig:6}(b) shows the minimum eigen-value of the density operator $\rho(t)$.
Negative eigen-values are basically obtained,
i.e., the positivity is violated in the non-Lindblad master equation
in Eq.~\eqref{eq:master_kappa_gamma_n}.
In this paper, we measure the positivity violation
by the absolute value of the minimum negative eigen-value.
At the early stage of Fig.~\ref{fig:6}(b), the positivity is gradually violated
together with the increase in the number of photons.
The violation is maximized before the number of photons starts to oscillate.
While the violation for $g = 1.0$ is more significant than for $g = 0.5$
at the early stage, the former starts to be suppressed earlier than the latter.
As the result, the positivity violation is most significant for $g = 0.5$
in the present demonstration.
Note that the positivity violation gradually diminishes afterward,
and it becomes negligible when the system reaches the ground state,
since the steady state $\orhoss = \ketg\brag$ guarantees
the minimum eigen-value equal to zero.

We next check the positivity in the dynamics
under the incoherent excitation ($n > 0$).
Figures \ref{fig:7}(a) and (b) show the development of the number of photons
and of the minimum eigen-value, respectively, in the case of $n = 0.1$.
For any $\rabi$, the number of photons finally becomes a larger value than in Fig.~\ref{fig:6}(a)
due to the incoherent excitation.
The results in the early stage are not strongly changed
from the dissipative situation in Fig.~\ref{fig:6}.
However, during the oscillation period, 
the positivity violation is clearly suppressed compared with Fig.~\ref{fig:6},
and it becomes negligible at an earlier time.
This is probably because the higher states have positive probabilities
thanks to the incoherent excitation, which diminishes the negative probability,
whereas the higher states have zero probability as $\orhoss = \ketg\brag$
in the dissipative situation.
We numerically checked that the positivity violation can be suppressed
by increasing $n$, i.e., by strengthening the incoherent excitation.

In conclusion, as far as we checked numerically,
the positivity can be violated
in the temporal development calculated by our non-Lindblad master equation,
although it is not violated when the system is close to the steady state.
The positivity violation becomes significant
when we start from a special initial state such as $\ketzz$,
while the violation remains small starting from $\ketg$.

\section{Advantage of non-Lindblad form} \label{sec:advantage}
We finally discuss the advantage of the non-Lindblad master equation
than the Lindblad one,
while the disadvantage is the violation of the positivity
as checked in the previous section.

One advantage is the agreement with the results by the MBCs
in the classical electrodynamics,
which was checked in Sec.~\ref{sec:compare}.
The Lindblad-type equation shows a discrepancy
due to the RWA to the SEC.

Another advantage is the simplicity in the calculation of the non-Lindblad-type
quantum Langevin equations (and stochastic differential ones
in Appendix \ref{app:SDE}).
As seen in Eq.~\eqref{eq:Langevin_non-Lindblad},
the non-Lindblad-type quantum Langevin equation
does not include the lowering and raising components
of system operators,
although they appear in the Lindblad-type equation,
Eq.~\eqref{eq:Langevin_Lindblad}.
We need the information of all the eigen-states of $\oHz$
for decomposing system operators to the lowering and raising components.
It is basically difficult to diagonalize $\oHz$
analytically and also numerically,
if the system of interest has much degrees of freedom.
The non-Lindblad-type quantum Langevin equations
does not require such a diagonalization,
and they are sometimes easily to be solved.
In the demonstration in this paper,
we can easily solve the non-Lindblad-type equations,
Eqs.~\eqref{eq:ABXY_nonLindblad}.
In contrast, the calculation by the Lindblad-type equations
are complicated as shown in Appendix \ref{app:eq_set}.
Basically, the non-Lindblad-type quantum Langevin equations
can be solved easier than the Lindblad-type ones,
since we do not need to diagonalize the Hamiltonian.
For example, in the study of the laser in the ultra-strong interaction regime,
the non-Lindblad-type quantum stochastic differential equations
enabled us to get easily the steady state solutions
\cite{Bamba2016Laser},
while the Hamiltonian of the laser system is hard to be diagonalized
due to the huge number of finite-level atoms interacting with the cavity modes.

\section{Summary} \label{sec:summary}
From the MBCs at an imperfect mirror of a Fabry-P\'erot cavity,
we derived a Hamiltonian connecting inside and outside the cavity
(SEC Hamiltonian)
in the good-cavity case \cite{Bamba2014SEC}.
From this Hamiltonian,
in the Born-Markov approximation but without the RWA to the SEC Hamiltonian,
the master equation of a non-Lindblad form is derived
even for frequency-independent loss rates $\kappa_j$ of the cavity modes.
Then, the positivity of the density operator is not guaranteed.
For transforming it to the Lindblad form,
we need to apply the RWA to the SEC Hamiltonian.
However, we found that the absorption spectra by the non-Lindblad master equation
agree well with those by the reliable calculation by the MBCs,
while the Lindblad master equation shows a discrepancy from them
in the ultra-strong light-matter interaction regime
with a large broadening (high damping rate $\gamma$).
In this way,
in the similar manner as in the studies of quantum Brownian motion
\cite{Caldeira1983PA,Grabert1988PR,Diosi1995EL,Munro1996PRA,gardiner04,Barnett2005PRA,Breuer2006},
for the consistency with the physical laws
(Maxwell equations or MBCs),
we sometimes need to consider a non-Lindblad master equation
derived without the RWA to the SEC,
while the mathematical requirement (positivity of density operator)
is not guaranteed.

From the viewpoint of studying the ultra-strong light-matter interaction regime,
it still remains unclear how the response, dissipation, and noise
are changed before and after the super-radiant phase transition (SRPT)
\cite{Mallory1969PR,Hepp1973AP,Wang1973PRA}
both for the non-equilibrium analogue \cite{Baumann2010N,Baumann2011PRL}
and for the original thermal-equilibrium SRPT,
which has not yet realized experimentally.
How to observe the SRPT is also open to dispute
\cite{Lolli2015PRL,Bamba2016circuitSRPT}.
Since the spontaneous appearance of the coherent amplitude of photonic field
at the SRPT
corresponds to the appearance of a static electric or magnetic field
in the system,
the physics involving the SEC is expected to be changed strongly
reflecting the MBCs between the environment
and the system with the static field.
For the correct investigation of the response, dissipation, and noise
after the SRPT, the consistency with the MBCs is essential
as we pursued in the present paper.

\begin{acknowledgments}
This work was funded by ImPACT Program of Council for Science, Technology and
Innovation (Cabinet Office, Government of Japan)
and by JSPS KAKENHI (Grants No.~26287087, No.~25247068, and No.~24-632).
\end{acknowledgments}

\appendix
\section{Derivation of master, quantum Langevin,
and quantum stochastic differential equations} \label{app:derive}
We will derive the master equation in App.~\ref{app:master}
and the quantum Langevin equation in App.~\ref{app:Langevin}
from the SEC Hamiltonian expressed as
\begin{equation} \label{eq:oHSEC_app} 
\oHSEC
= \int_0^{\infty}\dd\omega\
  \sum_j\ii\hbar\sqrt{\frac{\kappa_j(\omega)}{2\pi}} \oAA_j \left[ \ofd(\omega) - \of(\omega) \right].
\end{equation}
As will be discussed in Sec.~\ref{app:SDE},
the quantum stochastic differential equation is also derived
from the master equation.
Whereas this SEC Hamiltonian has the same form as supposed in Sec.~\ref{sec:master_Langevin},
here we consider the $\omega$-dependent loss rate $\kappa(\omega)$
and a more general correlation of the environment as
\begin{subequations}
\begin{align}
\braket{\ofd(\omega)\of(\omega')} & = n(\omega)\delta(\omega-\omega'), \\
\braket{\of(\omega)\ofd(\omega')} & = [n(\omega)+1]\delta(\omega-\omega'),
\end{align}
\end{subequations}
where $n(\omega)$ is the expectation number of bosons with frequency $\omega$
in the environment.
The Hamiltonian of the environment is
\begin{equation}
\oHenv = \int_0^{\infty}\dd\omega\ \hbar\omega\ofd(\omega)\of(\omega).
\end{equation}
We here consider that the system of interest
described by $\oHz$ couples only with this environment
through $\oHSEC$ in Eq.~\eqref{eq:oHSEC_app}
without considering any other environments.
We will derive the Lindblad and non-Lindblad master equations,
and the former is derived by applying the RWA to
the SEC Hamiltonian in Eq.~\eqref{eq:oHSEC_app}
in the basis of the eigen-states $\{\ket{\mu}\}$ of $\oHz$ as
\begin{equation} \label{eq:oHSEC_RWA_app} 
\oHSEC^{\text{RWA}}
= \int_0^{\infty}\dd\omega\
\sum_j \ii\hbar\sqrt{\frac{\kappa_j(\omega)}{2\pi}}
  \left[ \ofd(\omega)\oAAp_j - \oAAm_j\of(\omega) \right],
\end{equation}
where the lowering and raising operators are defined as in Eqs.~\eqref{eq:lower_raise}.

\subsection{Master equation} \label{app:master}
Let us derive the master equation
in the similar manner as in Refs.~\cite{Breuer2006,gardiner04}.
In the interaction picture,
the equation of motion of density operator $\irhotot(t)$
of the whole system $\oH = \oHz + \oHSEC + \oHenv$ is written as
\begin{align}
\ddt{}\irhotot(t)
& = \frac{1}{\ii\hbar} \left[ \iHint(t), \irhotot(t_0) \right]
\nonumber \\ & \quad
  - \frac{1}{\hbar^2} \int_{t_0}^{t}\dd t'
    \left[ \iHint(t), \left[ \iHint(t'), \irhotot(t') \right] \right],
\label{eq:irhotot} 
\end{align}
where $t_0\rightarrow-\infty$ is the switch-on time of the SEC
and the operators are defined as
\begin{align}
\iHint(t) & \equiv \ee^{\ii(\oHz+\oHenv) t/\hbar} \oHint \ee^{-\ii(\oHz+\oHenv) t/\hbar}, \\
\irhotot(t) & \equiv \ee^{\ii(\oHz+\oHenv) t/\hbar} \orhotot(t) \ee^{-\ii(\oHz+\oHenv) t/\hbar}.
\end{align}
Here, $\orhotot(t)$ is the density operator of the total system
in the Schr\"odinger picture.
We define the reduced density operator $\irho(t)$ of the system of interest
by taking the trace over the degrees of freedom in the environment as
\begin{equation}
\irho(t) \equiv \TrB[\irhotot(t)].
\end{equation}
Since we suppose that the fields in the environment has no coherence at the initial time $t_0$,
the first term in Eq.~\eqref{eq:irhotot} is
\begin{equation}
\TrB\left[ \iHint(t), \irhotot(t_0) \right] = 0.
\end{equation}
Then, we get the equation of motion of $\irho(t)$ as
\begin{equation} \label{eq:irho=} 
\ddt{}\irho(t)
= \int_{t_0}^{t} \frac{\dd t'}{-\hbar^2}\
    \TrB\left[ \iHint(t), \left[ \iHint(t'), \irhotot(t') \right] \right].
\end{equation}

Next, we use the Born approximation \cite{Breuer2006,gardiner04},
and the total density operator is represented as
\begin{equation}
\irhotot(t) = \irho(t) \otimes \irhoB.
\end{equation}
This is means that the environment is modified only slightly
and remains approximately in the initial state $\irhoB$,
which is justified
when the environment is huge enough and
the SEC is relatively weak.
Under the Born approximation, 
the integrand in Eq.~\eqref{eq:irho=} is rewritten as
\begin{widetext}
\begin{align} \label{eq:[H,[H,rho]]} 
& - \frac{1}{\hbar^2}\TrB\left[\iHint(t), \left[ \iHint(t'), \irhotot(t') \right]\right]
\nonumber \\
& = \sum_{j,j'} \int_0^{\infty}\dd\omega\ \frac{\sqrt{\kappa_j(\omega)\kappa_{j'}(\omega)}}{2\pi}
    \left\{
      [n(\omega)+1] \left(
        \ee^{\ii\omega(t-t')}
        \left[ \iAA_j(t), \irho(t') \iAA_{j'}(t') \right]
      + \ee^{-\ii\omega(t-t')}
        \left[ \iAA_{j'}(t')\irho(t'), \iAA_j(t) \right]
      \right)
\right. \nonumber \\ & \quad \left.
    + n(\omega) \left(
        \ee^{-\ii\omega(t-t')}
        \left[ \iAA_j(t), \irho(t') \iAA_{j'}(t') \right]
      + \ee^{\ii\omega(t-t')}
        \left[ \iAA_{j'}(t')\irho(t'), \iAA_j(t) \right]
      \right)
    \right\}.
\end{align}
On the other hand, when we perform the RWA to the SEC Hamiltonian,
we instead get
\begin{align}
& - \frac{1}{\hbar^2}\TrB\left[\iHint^{\text{RWA}}(t), \left[ \iHint^{\text{RWA}}(t'), \irhotot(t') \right]\right]
\nonumber \\
& = \sum_{j,j'} \int_0^{\infty}\dd\omega\ \frac{\sqrt{\kappa_j(\omega)\kappa_{j'}(\omega)}}{2\pi}
    \left\{
      [n(\omega)+1] \left(
        \ee^{\ii\omega(t-t')}
        \left[ \iAAp_j(t), \irho(t') \iAAm_{j'}(t') \right]
      + \ee^{-\ii\omega(t-t')}
        \left[ \iAAp_{j'}(t')\irho(t'), \iAAm_j(t) \right]
      \right)
\right. \nonumber \\ & \quad \left.
    + n(\omega) \left(
        \ee^{-\ii\omega(t-t')}
        \left[ \iAAm_j(t), \irho(t') \iAAp_{j'}(t') \right]
      + \ee^{\ii\omega(t-t')}
        \left[ \iAAm_{j'}(t')\irho(t'), \iAAp_j(t) \right]
      \right)
    \right\}.
\end{align}
Here, we use the Markov approximation
in the sense of Ref.~\cite{Breuer2006},
i.e., the reduced density operator $\irho(t')$ in the interaction picture
is replaced by $\irho(t)$
in a short enough coherence time of the environment.
Then, the equation of motion of the density operator $\orho(t)$
in the Schr\"odinger picture is obtained from Eqs.~\eqref{eq:irho=}
and \eqref{eq:[H,[H,rho]]} as
\begin{align}
\ddt{}\orho(t) & = \frac{1}{\ii\hbar}\left[\oHz, \orho\right]
  + \int_{t_0}^{t}\dd t'\ \sum_{j,j'}
    \int_0^{\infty}\dd\omega\
    \frac{\sqrt{\kappa_j(\omega)\kappa_{j'}(\omega)}}{2\pi}
\nonumber \\ & \times
    \left\{
      [n(\omega)+1] \left(
        \ee^{\ii\omega(t-t')}
        \left[ \oAA_j, \orho(t) \iAA_{j'}(t'-t) \right]
      + \ee^{-\ii\omega(t-t')}
        \left[ \iAA_{j'}(t'-t)\orho(t), \oAA_j \right]
      \right)
\right. \nonumber \\ & \quad \left.
    + n(\omega) \left(
        \ee^{-\ii\omega(t-t')}
        \left[ \oAA_j, \orho(t) \iAA_{j'}(t'-t) \right]
      + \ee^{\ii\omega(t-t')}
        \left[ \iAA_{j'}(t'-t)\orho(t), \oAA_j \right]
      \right)
    \right\},
\end{align}
where the density operator is defined as
\begin{equation}
\orho(t) \equiv \ee^{-\ii\oHz t/\hbar} \irho(t) \ee^{\ii\oHz t/\hbar}.
\end{equation}
Under the RWA to the SEC, we instead get
\begin{align}
\ddt{}\orho(t) & = \frac{1}{\ii\hbar}\left[\oHz, \orho\right]
  + \int_{t_0}^{t}\dd t'\ \sum_{j,j'}
    \int_0^{\infty}\dd\omega\
    \frac{\sqrt{\kappa_j(\omega)\kappa_{j'}(\omega)}}{2\pi}
\nonumber \\ & \times
    \left\{
      [n(\omega)+1] \left(
        \ee^{\ii\omega(t-t')}
        \left[ \oAAp_j, \orho(t) \iAAm_{j'}(t'-t) \right]
      + \ee^{-\ii\omega(t-t')}
        \left[ \iAAp_{j'}(t'-t)\orho(t), \oAAm_j \right]
      \right)
\right. \nonumber \\ & \quad \left.
    + n(\omega) \left(
        \ee^{-\ii\omega(t-t')}
        \left[ \oAAm_j, \orho(t) \iAAp_{j'}(t'-t) \right]
      + \ee^{\ii\omega(t-t')}
        \left[ \iAAm_{j'}(t'-t)\orho(t), \oAAp_j \right]
      \right)
    \right\}.
\end{align}
Here, the integral over $t'$ is rewritten for $t_0 \rightarrow -\infty$ as
\begin{subequations}
\begin{align}
\int_{-\infty}^t\dd t'\ \ee^{\mp\ii\omega(t-t')} \iAAp_{j'}(t'-t)
& = \sum_{\mu,\nu>\mu} \oAA_{j',\mu,\nu}
    \int_{-\infty}^t\dd t'\ \ee^{\ii(\omega_{\nu,\mu}\mp\omega)(t-t')}
\nonumber \\ &
  = \sum_{\mu,\nu>\mu} \oAA_{j',\mu,\nu} \left[
      \pi\delta(\omega_{\nu,\mu}\mp\omega)
    + \frac{\PP}{\ii(\omega_{\nu,\mu}\mp\omega)}
    \right], \\
\int_{-\infty}^t\dd t'\ \ee^{\pm\ii\omega(t-t')} \iAAm_{j'}(t'-t)
& = \sum_{\mu,\nu>\mu} \{\oAA_{j',\mu,\nu}\}^{\dagger}
    \int_{-\infty}^t\dd t'\ \ee^{-\ii(\omega_{\nu,\mu}\mp\omega)(t-t')} 
\nonumber \\ &
  = \sum_{\mu,\nu>\mu} \{\oAA_{j',\mu,\nu}\}^{\dagger} \left[
      \pi\delta(\omega_{\nu,\mu}\mp\omega)
    - \frac{\PP}{\ii(\omega_{\nu,\mu}\mp\omega)}
    \right].
\end{align}
\end{subequations}
The last terms contribute to the shift of the energies
due to the SEC (Lamb shift), and we neglect it in this paper.
Then, the master equation is obtained
without the RWA to the SEC as
\begin{align} \label{eq:master_woRWA} 
\ddt{}\orho(t)
& = \frac{1}{\ii\hbar}\left[\oHz, \orho\right]
  + \sum_{j,j'}\sum_{\mu,\nu>\mu}
    \frac{\sqrt{\kappa_j(\omega_{\nu,\mu})\kappa_{j'}(\omega_{\nu,\mu})}}{2}
    \left\{
      [n(\omega_{\nu,\mu})+1] \left(
        \left[ \oAA_j, \orho(t) \{\oAA_{j',\mu,\nu}\}^{\dagger} \right]
      + \left[ \oAA_{j',\mu,\nu}\orho(t), \oAA_j \right]
      \right)
\right. \nonumber \\ & \quad \left.
    + n(\omega_{\nu,\mu}) \left(
        \left[ \oAA_j, \orho(t) \oAA_{j',\mu,\nu} \right]
      + \left[ \{\oAA_{j',\mu,\nu}\}^{\dagger}\orho(t), \oAA_j \right]
      \right)
    \right\}.
\end{align}
For $\omega$-independent $\kappa$ and $n(\omega) = 0$,
this master equation is reduced to the non-Lindblad master equation
in Eq.~\eqref{eq:master_non-Lindblad_jj}.
Further, when the system of interest couples with the environment
only through an operator $\oAA$,
it is reduced to Eq.~\eqref{eq:master_non-Lindblad}.
When $n(\omega) \neq 0$ and the system of interest couples with the environment
only through an operator $\oAA$,
the above master equation is reduced to
Eq.~\eqref{eq:master_non-Lindblad_w_n}
for $\omega$-dependent $\kappa(\omega)$
and to Eq.~\eqref{eq:master_non-Lindblad_n}
for $\omega$-independent $\kappa$.
On the other hand, under the RWA to the SEC, we get
\begin{align} \label{eq:master_wRWA} 
\ddt{}\orho(t)
& = \frac{1}{\ii\hbar}\left[\oHz, \orho\right]
  + \sum_{j,j'}\sum_{\mu,\nu>\mu}
    \frac{\sqrt{\kappa_j(\omega_{\nu,\mu})\kappa_{j'}(\omega_{\nu,\mu})}}{2}
    \left\{
      [n(\omega_{\nu,\mu})+1] \left(
        \left[ \oAAp_j, \orho(t) \{\oAA_{j',\mu,\nu}\}^{\dagger} \right]
      + \left[ \oAA_{j',\mu,\nu}\orho(t), \oAAm_j \right]
      \right)
\right. \nonumber \\ & \quad \left.
    + n(\omega_{\nu,\mu}) \left(
        \left[ \oAAm_j, \orho(t) \oAA_{j',\mu,\nu} \right]
      + \left[ \{\oAA_{j',\mu,\nu}\}^{\dagger}\orho(t), \oAAp_j \right]
      \right)
    \right\}.
\end{align}
\end{widetext}
For $\omega$-independent $\kappa$ and $n(\omega) = 0$,
this master equation is reduced to the Lindblad master equation
in Eq.~\eqref{eq:master_Lindblad_jj}.
Further, when the system of interest couples with the environment
only through an operator $\oAA$,
it is reduced to Eq.~\eqref{eq:master_s}.
When $n(\omega) \neq 0$ and the system of interest couples with the environment
only through an operator $\oAA$,
the above master equation is reduced to
Eq.~\eqref{eq:master_s_w_n} for $\omega$-dependent $\kappa(\omega)$
and to Eq.~\eqref{eq:master_s_n}
for $\omega$-independent $\kappa$.

\subsection{Quantum Langevin equation} \label{app:Langevin}
Let us next derive the quantum Langevin equation
in the similar manner as in Ref.~\cite{gardiner04}.
From the SEC Hamiltonian without the RWA in Eq.~\eqref{eq:oHSEC_app},
the Heisenberg equation of $\of(\omega)$ is derived as
\begin{equation}
\ddt{}\of(\omega,t) = -\ii\omega \of(\omega,t)
+ \sum_j \sqrt{\frac{\kappa_j(\omega)}{2\pi}} \oAA_j(t).
\end{equation}
This equation is rewritten for $t_0\rightarrow-\infty$ as
\begin{subequations}
\begin{align}
\of(\omega,t)
& = \ee^{-\ii\omega(t-t_0)} \of(\omega,t_0)
\nonumber \\ & \quad
  + \sum_j \sqrt{\frac{\kappa_j(\omega)}{2\pi}}
    \int_{t_0}^{t}\dd t\ \ee^{-\ii\omega(t-t')} \oAA_j(t') \\
& = \ee^{-\ii\omega(t-t_0)} \of(\omega,t_0)
  + \sum_j \sqrt{\frac{\kappa_j(\omega)}{2\pi}}
    \int_{-\infty}^{\infty}\dd\omega'\
\nonumber \\ & \quad \times
    \ee^{-\ii\omega't}
    \left[ \pi\delta(\omega-\omega') - \frac{\ii\PP}{\omega-\omega'} \right]
    \oAA_j(\omega').
\end{align}
\end{subequations}
The last term also contribute to the Lamb shift,
and we neglect it in this paper.
Then, the quantum Langevin equation
without the RWA to the SEC is obtained
from the Heisenberg equation for arbitrary operator $\oS$
of the system of interest as
\begin{align}
\ddt{}\oS
& = \frac{1}{\ii\hbar}\left[ \oS, \oH_0 \right]
\nonumber \\ & \quad
  -  \int_0^{\infty}\dd\omega \left\{ \sum_j \left[ \oS, \oAA_j \right]
    \sqrt{\frac{\kappa_j(\omega)}{2\pi}} \of(\omega,t) + \Hc \right\} \\
& \approx \frac{1}{\ii\hbar}\left[ \oS, \oH_0 \right]
- \int_0^{\infty}\dd\omega\ \left\{
  \sum_j \left[ \oS, \oAA_j \right] \ee^{-\ii\omega t}\sqrt{\kappa_j(\omega)}
\right. \nonumber \\ & \quad \left. \times
  \left[ \oain(\omega) + \sum_{j'}\frac{\sqrt{\kappa_{j'}(\omega)}}{2}\oAA_{j'}(\omega) \right]
  + \Hc
  \right\},
\end{align}
where the input operator is defined as
\begin{equation}
\oain(\omega) = \ee^{\ii\omega t_0} \of(\omega,t_0) / \sqrt{2\pi}.
\end{equation}
This satisfies $[\oain(\omega), \oaind(\omega')] = \delta(\omega-\omega')/2\pi$.
The input-output relation is obtained as \cite{gardiner04,walls08}
\begin{equation}
\oaout(\omega) = \oain(\omega) + \sum_j \sqrt{\kappa_j(\omega)}\oAA_j(\omega).
\end{equation}
On the other hand, under the RWA to the SEC,
the quantum Langevin equation is derived
from the SEC Hamiltonian in Eq.~\eqref{eq:oHSEC_RWA_app} as
\begin{align}
\ddt{}\oS
& \approx \frac{1}{\ii\hbar}\left[ \oS, \oH_0 \right]
- \int_0^{\infty}\dd\omega\ \left\{
  \sum_j \left[ \oS, \oAAm_j \right] \ee^{-\ii\omega t}\sqrt{\kappa_j(\omega)}
\right. \nonumber \\ & \quad \left. \times
  \left[ \oain(\omega) + \sum_{j'}\frac{\sqrt{\kappa_{j'}(\omega)}}{2}\oAAp_{j'}(\omega) \right]
  + \Hc
  \right\}.
\end{align}
The input-output relation is obtained as
\begin{equation}
\oaout(\omega) = \oain(\omega) + \sum_j \sqrt{\kappa_j(\omega)}\oAAp_j(\omega).
\end{equation}

\subsection{Quantum stochastic differential equation} \label{app:SDE}
From the master equation
in Eq.~\eqref{eq:master_woRWA}
derived without the RWA to the SEC,
in the case of the $\omega$-independent $\kappa$
and flat distribution $n(\omega) = n$,
the corresponding quantum stochastic differential equation
in Itoh's form \cite{gardiner04}
is obtained for arbitrary operator $\oS$ of the system of interest as
\begin{align} \label{eq:QSDE_partial_preRWA} 
\dd\oS
& = \frac{1}{\ii\hbar}\left[ \oS, \oHz \right]\dd t
\nonumber \\ & \quad
+ \sum_{j,j'}\frac{\sqrt{\kappa_j\kappa_{j'}}}{2}(n+1)
  \left\{\oAAm_{j'}[\oS, \oAA_j]+[\oAA_j, \oS]\oAAp_{j'}\right\}\dd t
\nonumber \\ & \quad
+ \sum_{j,j'}\frac{\sqrt{\kappa_j\kappa_{j'}}}{2} n
  \left\{\oAAp_{j'}[\oS, \oAA_j]+[\oAA_j, \oS]\oAAm_{j'}\right\}\dd t
\nonumber \\ & \quad
- \sum_j\sqrt{\kappa_j}\left\{ [ \oS, \oAA_j ] \dd\oF(t)
  + \dd\oFd(t)[ \oAA_j, \oS] \right\},
\end{align}
where the fluctuation operator satisfies
\begin{subequations}
\begin{align}
\dd\oF(t)^2 = \dd\oFd(t)^2 & = 0, \\
\dd\oFd(t)\dd\oF(t) & = n \dd t, \\
\dd\oF(t)\dd\oFd(t) & = (n+1) \dd t.
\end{align}
\end{subequations}
On the other hand,
from the master equation in Eq.~\eqref{eq:master_wRWA}
derived under the RWA to the SEC, we get
\begin{align}
\dd\oS
& = \frac{1}{\ii\hbar}\left[ \oS, \oHz \right]\dd t
\nonumber \\ & \quad
+ \sum_{j,j'}\frac{\sqrt{\kappa_j\kappa_{j'}}}{2}(n+1)
  \left\{\oAAm_{j'}[\oS, \oAAp_j]+[\oAAm_j, \oS]\oAAp_{j'}\right\}\dd t
\nonumber \\ & \quad
+ \sum_{j,j'}\frac{\sqrt{\kappa_j\kappa_{j'}}}{2} n
  \left\{\oAAp_{j'}[\oS, \oAAm_j]+[\oAAp_j, \oS]\oAAm_{j'}\right\}\dd t
\nonumber \\ & \quad
- \sum_j\sqrt{\kappa_j}\left\{ [ \oS, \oAAm_j ] \dd\oF(t)
  + \dd\oFd(t)[ \oAAp_j, \oS] \right\}.
\end{align}
Replacing $\oAAp_{j}$ and $\oAAm_{j}$ by $\hat{c}$ and $\hat{c}^{\dagger}$
($\oAA_j = \hat{c} + \hat{c}^{\dagger}$),
respectively, this equation is certainly reduced
to the Itoh's quantum stochastic differential equation
shown in Ref.~\cite{gardiner04}.

\section{Non-Lindblad master equations for $\omega$-dependent $\kappa(\omega)$}\label{app:master_n}
{
\tabcolsep = .5em
\begin{table*}[tbp]
\caption{Validity of six types of master equations
both for $\omega$-independent cavity loss rate $\kappa$ and
distribution $n$ of environment
and for $\omega$-deponent $\kappa(\omega)$ or $n(\omega)$.}
\label{tab:A1}
\begin{tabular}{lll||lll|lll}
&&& \multicolumn{3}{c|}{$\omega$-independent $\kappa$ and $n$} &
\multicolumn{3}{c}{$\omega$-dependent $\kappa(\omega)$ or $n(\omega)$} \\
&RWA to SEC&& Weak & Strong & Ultra-strong & Weak & Strong & Ultra-strong \\ \hline \hline
Eq.~\eqref{eq:master_a_n} or \eqref{eq:master_a} & Photon-based & Lindblad & Good & Good & Bad & Good & Bad & Bad \\
Eq.~\eqref{eq:master_s_n} or \eqref{eq:master_s} & Eigen-state-based & Lindblad & Good & Good & $^*$Good & Good & Bad & Bad \\
Eq.~\eqref{eq:master_non-Lindblad_n} or \eqref{eq:master_non-Lindblad} & No & Non-Lindblad & Good & Good & Good & Good & Bad & Bad \\
\hline
Eq.~\eqref{eq:master_s_w_n} & Eigen-state-based & Non-Lindblad & Good & Good & $^*$Good & Good & Good & $^*$Good \\
Eq.~\eqref{eq:master_s_w_post_n} & Eigen-state-based & Lindblad & Bad & Good & $^*$Good & Bad & Good & $^*$Good \\
Eq.~\eqref{eq:master_non-Lindblad_w_n} & No & Non-Lindblad & Good & Good & Good & Good & Good & Good \\
\end{tabular}
\\$^*$Good quantitatively for narrow enough broadening avoiding mode overlaps.
\end{table*}
}
Whereas we basically supposed the zero-temperature environment
in the main text,
we can in general consider that
the environment has a population distribution as
$\braket{\ofd(\omega)\of(\omega')} = \delta(\omega-\omega')n(\omega)$.
For $\omega$-independent $\kappa$ and $n$ in the frequency range of interest,
the master equations in
Eqs.~\eqref{eq:master_a}, \eqref{eq:master_s}, and \eqref{eq:master_non-Lindblad}
are rewritten, respectively, as
\begin{align} \label{eq:master_a_n} 
\ddt{}\orho
& = \oLLz[\orho]
  + \frac{\kappa}{2} (n+1) \left(
        \left[ \oa, \orho \oad \right]
      + \left[ \oa\orho, \oad \right]
      \right)
\nonumber \\ & \quad
  + \frac{\kappa}{2} n \left(
        \left[ \oad, \orho \oa \right]
      + \left[ \oad\orho, \oa \right]
      \right),
\end{align}
\begin{align} \label{eq:master_s_n} 
\ddt{}\orho
& = \oLLz[\orho]
  + \frac{\kappa}{2} (n+1) \left(
        \left[ \oAAp, \orho \oAAm \right]
      + \left[ \oAAp\orho, \oAAm \right]
      \right)
\nonumber \\ & \quad
  + \frac{\kappa}{2} n \left(
        \left[ \oAAm, \orho \oAAp \right]
      + \left[ \oAAm\orho, \oAAp \right]
      \right),
\end{align}
\begin{align} \label{eq:master_non-Lindblad_n} 
\ddt{}\orho
& = \oLLz[\orho]
  + \frac{\kappa}{2} (n+1) \left(
        \left[ \oAA, \orho \oAAm \right]
      + \left[ \oAAp\orho, \oAA \right]
      \right)
\nonumber \\ & \quad
    + \frac{\kappa}{2} n \left(
        \left[ \oAA, \orho \oAAp \right]
      + \left[ \oAAm\orho, \oAA \right]
      \right).
\end{align}
The validity of these master equations is re-summarized in Tab.~\ref{tab:A1}.

When the cavity loss $\kappa(\omega)$ and distribution $n(\omega)$
depends on the frequency $\omega$ in the frequency range of interest,
the above three master equations are not appropriate in general.
However,
if the frequency range of interest is only around the cavity resonance $\wc$
and $\kappa(\omega)$ is not strongly varied in it,
i.e., in the weak light-matter interaction regime,
the above three master equations are basically valid.
In this sense, in Tab.~\ref{tab:A1}, we denote that they are Good
in the weak interaction regime even for $\omega$-dependent $\kappa(\omega)$ or $n(\omega)$.

In the normally strong or ultra-strong light-matter interaction regime
for $\omega$-dependent $\kappa(\omega)$ or $n(\omega)$,
as derived in App.~\ref{app:derive},
the master equations are never expressed with the photon operator $\oa$
as in Eq.~\eqref{eq:master_a_n},
but they are represented as similar as the two master equations
in Eqs.~\eqref{eq:master_s_n} and \eqref{eq:master_non-Lindblad_n}.
Instead of Eq.~\eqref{eq:master_s_n}
derived under the RWA to the SEC Hamiltonian,
the master equation for $\omega$-dependent $\kappa(\omega)$ and $n(\omega)$ is obtained in the Born-Markov approximation
(we obey Ref.~\cite{Breuer2006} concerning the meaning of the Markov approximation
as performed in App.~\ref{app:derive} of this paper) as
\begin{align} \label{eq:master_s_w_n} 
\ddt{}\orho
& = \oLLz[\orho]
  + \sum_{\mu,\nu>\mu}
    \frac{\kappa(\omega_{\nu,\mu})}{2}
\nonumber \\ & \quad \times
    \left\{
      [n(\omega_{\nu,\mu})+1] \left(
        \left[ \oAAp, \orho \{\oAA_{\mu,\nu}\}^{\dagger} \right]
      + \left[ \oAA_{\mu,\nu}\orho, \oAAm \right]
      \right)
\right. \nonumber \\ & \quad \left.
    + n(\omega_{\nu,\mu}) \left(
        \left[ \oAAm, \orho \oAA_{\mu,\nu} \right]
      + \left[ \{\oAA_{\mu,\nu}\}^{\dagger}\orho, \oAAp \right]
      \right)
    \right\}.
\end{align}
Whereas, under the RWA to the light-matter interaction,
$\oAAp$ and $\oAAm$ are reduced to $\oa$ and $\oad$, respectively,
we cannot rewrite the summation of $\kappa(\omega_{\nu,\mu})\oAA_{\mu,\nu}$
[$\times n(\omega_{\nu,\mu})$] and its Hermite conjugate
simply by the photon annihilation or creation operators.
In this way, for the $\omega$-dependent $\kappa(\omega)$ or $n(\omega)$,
we need to use the master equation in Eq.~\eqref{eq:master_s_w_n}
even in the normally strong light-matter interaction regime,
instead of Eq.~\eqref{eq:master_a_n} or Eq.~\eqref{eq:master_s_n}.

Note that, while the two master equations \eqref{eq:master_a_n} and \eqref{eq:master_s_n}
are of the Lindblad form,
Eq.~\eqref{eq:master_s_w_n} is of a non-Lindblad form.
In order to transform it to the Lindblad form,
we sometimes neglect rapidly oscillating terms
(called the post-trace RWA in Ref.~\cite{Fleming2010JPA}
and the RWA in Eqs.~\eqref{eq:RWA_eigen} and \eqref{eq:RWA_photon}
are called the pre-trace RWA),
and Eq.~\eqref{eq:master_s_w_n} is approximated as
\begin{align} \label{eq:master_s_w_post_n} 
\ddt{}\orho
& = \oLLz[\orho]
  + \sum_{\mu} \sum_{\nu>\mu}
    \frac{\kappa(\omega_{\nu,\mu})}{2}
    \Big\{
      [n(\omega_{\nu,\mu})+1]
\nonumber \\ & \quad \times
      \left(
        \left[ \oAA_{\mu,\nu}, \orho \{\oAA_{\mu,\nu}\}^{\dagger} \right]
      + \left[ \oAA_{\mu,\nu}\orho, \{\oAA_{\mu,\nu}\}^{\dagger} \right]
      \right)
\nonumber \\ & \quad
    + n(\omega_{\nu,\mu}) \left(
        \left[ \{\oAA_{\mu,\nu}\}^{\dagger}, \orho \oAA_{\mu,\nu} \right]
      + \left[ \{\oAA_{\mu,\nu}\}^{\dagger}\orho, \oAA_{\mu,\nu} \right]
      \right)
    \Big\}.
\end{align}
This is certainly of the Lindblad form.
However, this approximation is valid only
when the transitions between the eigen-states are well identified
under strong enough light-matter interaction.
Then, the master equation in Eq.~\eqref{eq:master_s_w_post_n}
is not appropriate in the weak interaction regime
(we note this fact in Tab.~\ref{tab:A1}).
In such case, we should rather use the simple Lindblad master equation
in Eq.~\eqref{eq:master_a_n}
or the non-Lindblad master equation in Eq.~\eqref{eq:master_s_w_n}.
In this way,
we have faced the problem of the Lindblad form
(positivity of density operator)
in the study of cavity QED
even in the weak and normally strong interaction regimes.

For $\omega$-dependent $\kappa(\omega)$ or $n(\omega)$,
instead of Eq.~\eqref{eq:master_non-Lindblad_n},
the master equation is derived
in the Born-Markov approximation but without the RWA to the SEC as
\begin{align} \label{eq:master_non-Lindblad_w_n} 
\ddt{}\orho
& = \oLLz[\orho]
  + \sum_{\mu} \sum_{\nu>\mu}
    \frac{\kappa(\omega_{\nu,\mu})}{2}
\nonumber \\ & \quad \times
    \left\{ [n(\omega_{\nu,\mu})+1]
      \left(
        \left[ \oAA, \orho \{\oAA_{\mu,\nu}\}^{\dagger} \right]
      + \left[ \oAA_{\mu,\nu}\orho, \oAA \right]
      \right)
\right. \nonumber \\ & \quad \left. 
    + n(\omega_{\nu,\mu})
      \left(
        \left[ \oAA, \orho \oAA_{\mu,\nu} \right]
      + \left[ \{\oAA_{\mu,\nu}\}^{\dagger}\orho, \oAA \right]
      \right)
    \right\}.
\end{align}
Compared with Eq.~\eqref{eq:master_s_w_n},
the lowering and raising components $\oAAp$ and $\oAAm$ are replaced by the original operator $\oAA$,
while both of them are of the non-Lindblad form.

In this way, there are two kinds of the non-Lindblad forms:
One is found in Eq.~\eqref{eq:master_s_w_n},
which appears also in the weak and normally strong light-matter interaction regimes
in the case of $\omega$-dependent $\kappa(\omega)$ or $n(\omega)$.
The other is found in Eqs.~\eqref{eq:master_non-Lindblad_n} and \eqref{eq:master_non-Lindblad},
whose difference from the Lindblad master equation
is highlighted in the ultra-strong light-matter interaction regime
with a large broadening.
In order to analyze the latter non-Lindblad form,
we considered the $\omega$-independent $\kappa$ with $n(\omega) = 0$
in the main text.
Eq.~\eqref{eq:master_non-Lindblad_w_n} contains
both the two non-Lindblad contributions.

Note that,
if all the environments have the same temperature $T$
and bosonic environments show the Bose distribution $n(\omega) = 1/(\ee^{\hbar\omega/\kB T}-1)$,
the thermal state $\rho = \ee^{-\oHz/\kB T}$
is obtained as a steady state of the master equations
in Eqs.~\eqref{eq:master_s_w_n}, \eqref{eq:master_s_w_post_n},
and \eqref{eq:master_non-Lindblad_w_n}.
On the other hand,
if all the environments are at zero temperature,
the steady state is guaranteed as the ground state of $\oHz$
in the master equations in Eqs.~\eqref{eq:master_s}, \eqref{eq:master_non-Lindblad},
although Eq.~\eqref{eq:master_a} does not gives the ground state
of the Hamiltonian such as in Eq.~\eqref{eq:oHz_simple}
(in ultra-strong interaction regime) \cite{Beaudoin2011PRA}.

\section{Calculation of absorption by Lindblad-type equations} \label{app:eq_set}
Whereas the absorption can be calculated by the simple equation set
for the non-Lindblad-type quantum Langevin equations
as discussed in Sec.~\ref{sec:by_non-Lindblad},
here we show the calculation method
applicable to both the Lindblad- and non-Lindblad-type equations
not only for the cavity loss but also for the excitation damping.

We first define the array of operators for wavenumber $k_j$ as
\begin{equation}
\ovv_j = \begin{pmatrix} \oAA_j & \oBB_j & \oXX_j & \oYY_j \end{pmatrix}^{\text{T}},
\end{equation}
where $\text{T}$ means the matrix transpose.
For both the Lindblad- and non-Lindblad-type quantum Langevin equations,
the equation set for the positive-frequency components is expressed as
\begin{align} \label{eq:Langevin_w} 
&
(\mM^0_{j}+\ii\omega\munit) \ovv_j(\omega)
\nonumber \\ &
= \sqrt{\kappa_j}\mMk_j
    \left[ \sum_{j'} (\sqrt{\kappa_{j'}}/2)\oAA_{j'}(\omega) + \oain(\omega) \right]
\nonumber \\ & \quad
  + \mMg_j\left[ (\gamma/2)\oXX_j(\omega) + \sqrt{\gamma}\obin_j(\omega) \right].
\end{align}
Here, the matrix on the left-hand side is derived
from $\oHz^v$ in the velocity form as
\begin{equation}
\mM^0_j = \begin{pmatrix}
0 & -ck & 0 & 0 \\
ck+4\rabiC_k{}^2/\wa & 0 & 0 & 2\rabiC_{k} \\
-2\rabiC_k & 0 & 0 & -\wa \\
0 & 0 & \wa & 0
\end{pmatrix},
\end{equation}
and from $\oHz^r$ in the length form as
\begin{equation}
\mM^0_j = \begin{pmatrix}
0 & -ck & 2\rabiP_k & 0 \\
ck & 0 & 0 & 0 \\
0 & 0 & 0 & -\wa \\
0 & -2\rabiP_k & \wa+4\rabia^2\wa & 0
\end{pmatrix}.
\end{equation}
We checked numerically that the same absorption spectra
are obtained in the two form.
The coefficient vectors on the right-hand side
are expressed as
\begin{equation}
\mMk_j = \begin{cases}
\begin{pmatrix} 0 & \ii2 & 0 & 0 \end{pmatrix}^{\text{T}} & \text{non-Lindblad}\\
\sum_{\zeta}
\begin{pmatrix} Q_{j,\zeta} & \varPi_{j,\zeta} & X_{j,\zeta} & Y_{j,\zeta} \end{pmatrix}^{\text{T}}
Q_{j,\zeta}^*
& \text{Lindblad}
\end{cases}
\end{equation}
\begin{equation}
\mMg_j = \begin{cases}
\begin{pmatrix} 0 & 0 & 0 & \ii2 \end{pmatrix}^{\text{T}} & \text{non-Lindblad}\\
\sum_{\zeta}
\begin{pmatrix} Q_{j,\zeta} & \varPi_{j,\zeta} & X_{j,\zeta} & Y_{j,\zeta} \end{pmatrix}^{\text{T}}
X_{j,\zeta}^*
& \text{Lindblad}
\end{cases}
\end{equation}
where we defined
\begin{subequations}
\begin{align}
Q_{j,\zeta} & = (w_{j,\zeta}^*-y_{j,\zeta}^*), \\
\varPi_{j,\zeta} & = \ii(w_{j,\zeta}^*+y_{j,\zeta}^*), \\
X_{j,\zeta} & = (x_{j,\zeta}^*-z_{j,\zeta}^*), \\
Y_{j,\zeta} & = \ii(x_{j,\zeta}^*+z_{j,\zeta}^*).
\end{align}
\end{subequations}
$\mMk_j$ and $\mMg_j$ governs the SEC of the cavity loss
and excitation damping, respectively,
reflecting whether they are treated in Lindblad- or non-Lindblad-form.
The equation set is rewritten as
\begin{equation} \label{eq:Mv=Ca+Cb} 
\sum_{j'}\mM_{j,j'}(\omega) \ovv_{j'}(\omega)
= \sqrt{\kappa_j} \mMk_j \oain(\omega)
+ \sqrt{\gamma} \mMg_j \obin_j(\omega),
\end{equation}
where the coefficient matrix is expressed as
\begin{align}
\mM_{j,j'}(\omega)
& = \delta_{j,j'}\left[ \mM_0 - \frac{\gamma}{2}\mMg_j
  \times\begin{pmatrix} 0 & 0 & 1 & 0 \end{pmatrix}
+ \ii\omega\bm{1}\right]
\nonumber \\ & \quad
 - \frac{\sqrt{\kappa_j\kappa_{j'}}}{2}\mMk_j
   \times\begin{pmatrix} 1 & 0 & 0 & 0 \end{pmatrix}.
\end{align}
From the input-output relation,
we can get the reflection coefficient $r(\omega)$ as
\begin{equation}
\oaout(\omega) = \oain(\omega) + \sum_j \sqrt{\kappa_j}\oAA_j(\omega)
= r(\omega) \oain(\omega) + \ldots
\end{equation}
Then, we can calculate the reflection and absorption spectra.

For deriving Eq.~\eqref{eq:Langevin_w},
we in fact replaced all the lowering operators
with the original ones such as $\oAAp_j(\omega)$ with $\oAA_j(\omega)$ for $\omega > 0$.
On the other hand, if we treat both the cavity loss and damping
in the Lindblad-type treatment,
we can get simple equation set for polariton annihilation operators $\{\op_{j,\zeta}\}$
by inversely replacing $\oAA_j(\omega)$ with $\oAAp_j(\omega)$ as
\begin{align}& \label{eq:iwp=Q+X} 
\ii(\omega-\omega_{j,\zeta})\op_{j,\zeta}(\omega)
\nonumber \\
& = Q_{j,\zeta}^*\sqrt{\kappa_j}\left[\sum_{j'}\frac{\sqrt{\kappa_{j'}}}{2}\sum_{\zeta'}Q_{j',\zeta'}\op_{j',\zeta'}(\omega) + \oain(\omega)\right]
\nonumber \\ & \quad
+ X_{j,\zeta}^*\left[\frac{\gamma}{2}\sum_{\zeta'}X_{j,\zeta'}\op_{j,\zeta'}(\omega) + \sqrt{\gamma}\obin_j(\omega)\right].
\end{align}
From this equation set and the input-output relation
\begin{equation}
\oaout(\omega) = \oain(\omega) + \sum_j \sqrt{\kappa_j}\sum_{\zeta}Q_{j,\zeta}\op_{j,\zeta},
\end{equation}
we can also calculate the reflection coefficient $r(\omega)$,
reflection and absorption spectra.
As far as we checked numerically,
we get approximately the same absorption spectra
by Eq.~\eqref{eq:iwp=Q+X} and by Eq.~\eqref{eq:Mv=Ca+Cb}
with the Lindblad-type treatment for both the cavity loss and damping.
In this way, $\oAA_j(\omega)$ and $\oAAp_j(\omega)$ are approximately equivalent
in the parameter region of this paper.

\section{Other numerical results}\label{app:Lindblad_damping}
\begin{figure}[tbp]
\includegraphics[width=.9\linewidth]{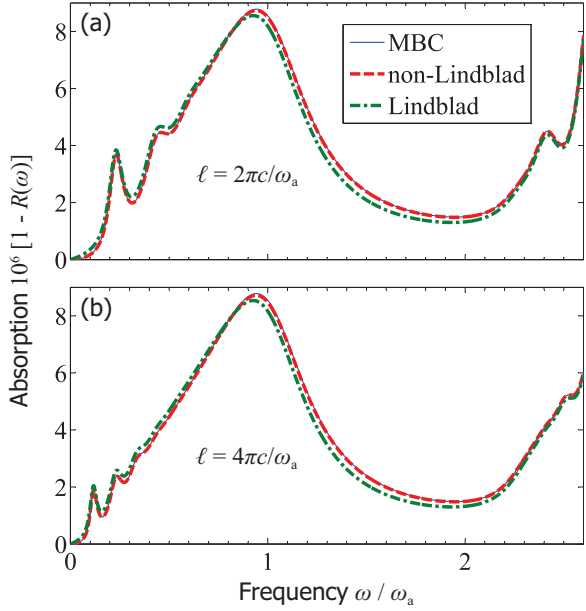}
\caption{Absorption spectra by three approaches:
by MBCs (blue solid line),
by non-Lindblad-type equation (red dashed line),
and by Lindblad-type equation (green dash-dotted line).
The cavity lengths are (a) $\cavlen = 2\pi c/\wa$
and (b) $\cavlen = 4\pi c/\wa$,
where the frequency of the lowest cavity mode is
$ck_1 = 0.5\wa$ and $0.25\wa$, respectively.
We basically get the same tendency
as in Fig.~\ref{fig:4}.
Parameters: $\rabi = 1$, $\gamma = 0.5\wa$, and $\varLambda_0 = 10^3$.
$k_j$ up to $j = 2000$ are considered.}
\label{fig:A1}
\end{figure}
In Fig.~\ref{fig:A1}, we show the absorption spectra
for (a) cavity length $\cavlen = 2\pi c/\wa$
and (b) $\cavlen = 4\pi c/\wa$,
where the frequency of the lowest cavity mode is
$ck_1 = 0.5\wa$ and $0.25\wa$, respectively.
The other parameters are $\rabi = 1$, $\gamma = 0.5\wa$, and $\varLambda = 10^3$.
The spectra by the three approaches are plotted with different lines.
We get basically the same tendency as the case of $\cavlen = \pi c/\wa$
discussed in the main text.

\begin{figure}[tbp]
\includegraphics[width=.9\linewidth]{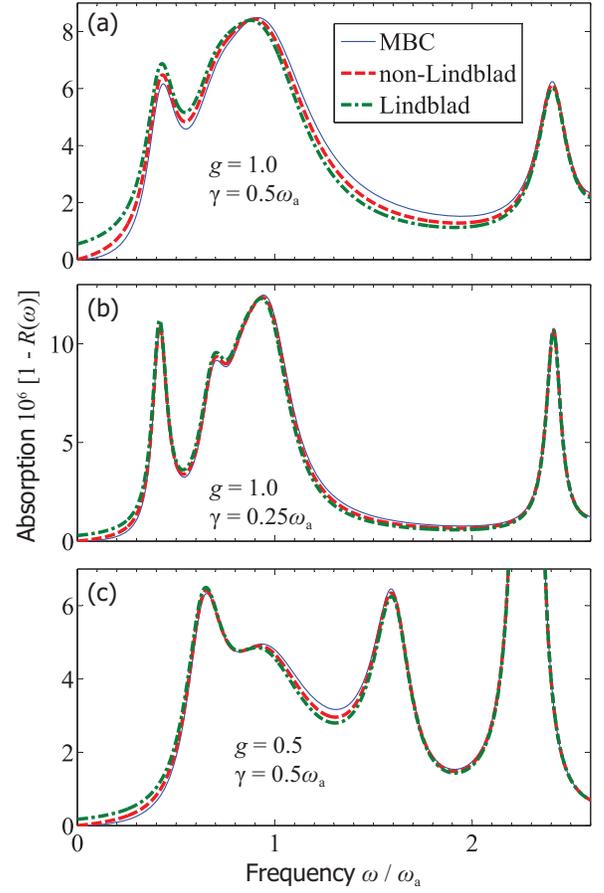}
\caption{Absorption spectra are calculated
with the Lindblad-type treatment for excitation damping.
For the cavity loss, we used the three approaches:
by the MBCs (blue solid line),
by non-Lindblad-type equation (red dashed line),
and by Lindblad-type equation (green dash-dotted line).
We considered (a) $\rabi = 1.0$ and $\gamma = 0.5\wa$,
(b) $\rabi = 1.0$ and $\gamma = 0.25\wa$, and
(c) $\rabi = 0.5$ and $\gamma = 0.5\wa$.
In contrast to the non-Lindblad-type treatment for damping
in Fig.~\ref{fig:4},
we can find a clear discrepancy between the spectra
by the MBCs and by the non-Lindblad-type equation
concerning the cavity loss.
However, for the Lindblad-type equation
concerning the cavity loss, we get a larger discrepancy,
and the tendency is similar as in Fig.~\ref{fig:4}
Parameters: $\cavlen = \pi c/\wa$ and $\varLambda_0 = 10^3$.
$k_j$ up to $j = 2000$ are considered.}
\label{fig:A2}
\end{figure}
In Fig.~\ref{fig:A2}, we show the absorption spectra
in the Lindblad-type treatment for the excitation damping.
The three curves are calculated by the MBCs (solid blue line),
quantum Langevin equation with the non-Lindblad-type treatment
for the cavity loss (red dashed line),
and that with the Lindblad-type treatment for the cavity loss
(green dash-dotted line).
The calculation method of the latter two are explained
in App.~\ref{app:eq_set}.
For the calculation by the MBCs,
we numerically calculated the dielectric function $\diep(\omega)$
of the medium in the Lindblad-type treatment for the excitation damping
as follows.
In the spatially infinite system as discussed in Sec.~\ref{sec:diep},
the quantum Langevin equation of the polariton annihilation operator
is obtained as
\begin{align}&
\ii(\omega-\omega_{k,\zeta})\op_{k,\zeta}(\omega)
\nonumber \\
& = X_{k,\zeta}^*\left[\frac{\gamma}{2}\sum_{\zeta'}X_{k,\zeta'}\op_{k,\zeta'}(\omega) + \sqrt{\gamma}\obin_k(\omega)\right].
\end{align}
The dispersion relation is obtained from
the zero determinant of the coefficient matrix as
\begin{equation}
\mathrm{det}
\begin{bmatrix}
\ii(\omega-\omega_{k,L}) - \frac{\gamma}{2}|X_{k,L}|^2&
- \frac{\gamma}{2}X_{k,L}^* X_{k,U} \\
- \frac{\gamma}{2}X_{k,U}^* X_{k,L} &
\ii(\omega-\omega_{k,U}) - \frac{\gamma}{2}|X_{k,U}|^2
\end{bmatrix}
= 0
\end{equation}
For given $\omega$, we numerically find a complex wavenumber $\kp(\omega)$
satisfying this equation
by using analytical expressions of $\omega_{k,\zeta}$ and $X_{k,\zeta}$.
Then, using this $\kp(\omega)$ and $\np(\omega) = c \kp(\omega) / \omega$,
the reflection coefficient $r(\omega)$ is calculated
by Eq.~\eqref{eq:r(w)}.

As seen in Fig.~\ref{fig:A2},
we get a clear discrepancy between the absorption spectra
by the MBCs and by the non-Lindblad-type equation,
while a larger discrepancy is obtained for the Lindblad-type equation
and these discrepancies are reduced
for smaller broadening as in Fig.~\ref{fig:A2}(b).
Although it is hard to catch correctly the reason of this new discrepancy
between the MBCs and the non-Lindblad-type equation,
it is rather natural because the two approaches are apparently different,
and the influences of the RWA to the SEC of the damping
are of course different in the two approaches.
Since we did not specify the mechanism of the damping,
we cannot determine which spectrum is correct
if the SEC Hamiltonian of damping is really expressed
such as in Eq.~\eqref{eq:RWA_eigen}
or the RWA to the SEC is justified by some reasons.
However, by looking the surprisingly good agreement
of the spectra by the MBCs and by the non-Lindblad-type equation
in Figs.~\ref{fig:4} and \ref{fig:A1},
we should basically not apply the RWA to the SEC
and use the SEC Hamiltonian such as in Eq.~\eqref{eq:oHSEC_simple}
for any SECs
in the ultra-strong light-matter interaction regime
with a large broadening.


\begin{thebibliography}{10}
\expandafter\ifx\csname href\endcsname\relax\def\href#1#2{#2}\fi

\bibitem{Senitzky1960PR}
I.~R. Senitzky, \href{http://link.aps.org/doi/10.1103/PhysRev.119.670}{Phys.
  Rev. {\bf 119}, 670 (1960)}.

\bibitem{Dekker1981PR}
H.~Dekker,
  \href{http://www.sciencedirect.com/science/article/pii/0370157381900338}{Physics
  Reports {\bf 80}, 1  (1981)}.

\bibitem{Caldeira1983PA}
A.~O. Caldeira and A.~J. Leggett,
  \href{http://www.sciencedirect.com/science/article/pii/0378437183900134}{Physica
  A {\bf 121}, 587  (1983)}.

\bibitem{Grabert1988PR}
H.~Grabert, P.~Schramm, and G.-L. Ingold,
  \href{http://www.sciencedirect.com/science/article/pii/0370157388900233}{Physics
  Reports {\bf 168}, 115 (1988)}.

\bibitem{Diosi1995EL}
L.~Di\'{o}si, \href{http://stacks.iop.org/0295-5075/30/i=2/a=001}{Europhys.
  Lett. {\bf 30}, 63 (1995)}.

\bibitem{Munro1996PRA}
W.~J. Munro and C.~W. Gardiner,
  \href{http://link.aps.org/doi/10.1103/PhysRevA.53.2633}{Phys. Rev. A {\bf
  53}, 2633 (1996)}.

\bibitem{gardiner04}
C.~W. Gardiner and P.~Zoller, {\em Quantum Noise: A Handbook Of Markovian And
  Non-markovian Quantum Stochastic Methods With Applications To Quantum
  Optics}, Springer Series in Synergetics (Springer-Verlag, Berlin, 2004),
  Third edition.

\bibitem{Barnett2005PRA}
S.~M. Barnett and J.~D. Cresser,
  \href{http://link.aps.org/doi/10.1103/PhysRevA.72.022107}{Phys. Rev. A {\bf
  72}, 022107 (2005)}.

\bibitem{Breuer2006}
H.-P. Breuer and F.~Petruccione, {\em The Theory of Open Quantum Systems},
  (Claredon Press, Oxford, 2006).

\bibitem{Lindblad1976CMP}
G.~Lindblad, \href{http://dx.doi.org/10.1007/BF01608499}{Comm.
  Math. Phys. {\bf 48}, 119 (1976)}.

\bibitem{walls08}
D.~F. Walls and G.~J. Milburn, {\em Quantum Optics},  (Springer-Verlag, Berlin,
  2008), 2nd edition.

\bibitem{Bamba2014SEC}
M.~Bamba and T.~Ogawa,
  \href{http://link.aps.org/doi/10.1103/PhysRevA.89.023817}{Phys. Rev. A
  {\bf 89}, 023817 (2014)}.

\bibitem{Ciuti2005PRB}
C.~Ciuti, G.~Bastard, and I.~Carusotto,
  \href{http://link.aps.org/doi/10.1103/PhysRevB.72.115303}{Phys. Rev. B {\bf
  72}, 115303 (2005)}.

\bibitem{Gunter2009N}
G.~Gunter, A.~A. Anappara, J.~Hees, A.~Sell, G.~Biasiol, L.~Sorba,
  S.~De~Liberato, C.~Ciuti, A.~Tredicucci, A.~Leitenstorfer, and R.~Huber,
  \href{http://dx.doi.org/10.1038/nature07838}{Nature {\bf 458}, 178 (2009)}.

\bibitem{Anappara2009PRB}
A.~A. Anappara, S.~De~Liberato, A.~Tredicucci, C.~Ciuti, G.~Biasiol, L.~Sorba,
  and F.~Beltram,
  \href{http://link.aps.org/doi/10.1103/PhysRevB.79.201303}{Phys. Rev. B {\bf
  79}, 201303 (2009)}.

\bibitem{Todorov2009PRL}
Y.~Todorov, A.~M. Andrews, I.~Sagnes, R.~Colombelli, P.~Klang, G.~Strasser, and
  C.~Sirtori,
  \href{http://link.aps.org/doi/10.1103/PhysRevLett.102.186402}{Phys. Rev.
  Lett. {\bf 102}, 186402 (2009)}.

\bibitem{Todorov2010PRL}
Y.~Todorov, A.~M. Andrews, R.~Colombelli, S.~De~Liberato, C.~Ciuti, P.~Klang,
  G.~Strasser, and C.~Sirtori,
  \href{http://link.aps.org/doi/10.1103/PhysRevLett.105.196402}{Phys. Rev.
  Lett. {\bf 105}, 196402 (2010)}.

\bibitem{Niemczyk2010NP}
T.~Niemczyk, F.~Deppe, H.~Huebl, E.~P. Menzel, F.~Hocke, M.~J. Schwarz, J.~J.
  Garcia-Ripoll, D.~Zueco, T.~Hummer, E.~Solano, A.~Marx, and R.~Gross,
  \href{http://dx.doi.org/10.1038/nphys1730}{Nat. Phys. {\bf 6}, 772 (2010)}.

\bibitem{Fedorov2010PRL}
A.~Fedorov, A.~K. Feofanov, P.~Macha, P.~Forn-D\'iaz, C.~J. P.~M. Harmans, and
  J.~E. Mooij,
  \href{http://link.aps.org/doi/10.1103/PhysRevLett.105.060503}{Phys. Rev.
  Lett. {\bf 105}, 060503 (2010)}.

\bibitem{Forn-Diaz2010PRL}
P.~Forn-D\'iaz, J.~Lisenfeld, D.~Marcos, J.~J. Garc\'ia-Ripoll, E.~Solano,
  C.~J. P.~M. Harmans, and J.~E. Mooij,
  \href{http://link.aps.org/doi/10.1103/PhysRevLett.105.237001}{Phys. Rev.
  Lett. {\bf 105}, 237001 (2010)}.

\bibitem{Schwartz2011PRL}
T.~Schwartz, J.~A. Hutchison, C.~Genet, and T.~W. Ebbesen,
  \href{http://link.aps.org/doi/10.1103/PhysRevLett.106.196405}{Phys. Rev.
  Lett. {\bf 106}, 196405 (2011)}.

\bibitem{Porer2012PRB}
M.~Porer, J.-M. M\'enard, A.~Leitenstorfer, R.~Huber, R.~Degl'Innocenti,
  S.~Zanotto, G.~Biasiol, L.~Sorba, and A.~Tredicucci,
  \href{http://link.aps.org/doi/10.1103/PhysRevB.85.081302}{Phys. Rev. B {\bf
  85}, 081302 (2012)}.

\bibitem{Scalari2012S}
G.~Scalari, C.~Maissen, D.~Tur\v{c}inkov\'a, D.~Hagenm\"uller, S.~De~Liberato,
  C.~Ciuti, C.~Reichl, D.~Schuh, W.~Wegscheider, M.~Beck, and J.~Faist,
  \href{http://www.sciencemag.org/content/335/6074/1323.abstract}{Science {\bf
  335}, 1323 (2012)}.

\bibitem{Zhang2014PRL}
Q.~Zhang, T.~Arikawa, E.~Kato, J.~L. Reno, W.~Pan, J.~D. Watson, M.~J. Manfra,
  M.~A. Zudov, M.~Tokman, M.~Erukhimova, A.~Belyanin, and J.~Kono,
  \href{http://link.aps.org/doi/10.1103/PhysRevLett.113.047601}{Phys. Rev.
  Lett. {\bf 113}, 047601 (2014)}.

\bibitem{Carmichael1973JPA}
H.~J. Carmichael and D.~F. Walls,
  \href{http://stacks.iop.org/0301-0015/6/i=10/a=014}{J. Phys. A {\bf 6}, 1552
  (1973)}.

\bibitem{Carmichael1974PRA}
H.~J. Carmichael and D.~F. Walls,
  \href{http://link.aps.org/doi/10.1103/PhysRevA.9.2686}{Phys. Rev. A {\bf 9},
  2686 (1974)}.

\bibitem{Scala2007PRA}
M.~Scala, B.~Militello, A.~Messina, J.~Piilo, and S.~Maniscalco,
  \href{http://link.aps.org/doi/10.1103/PhysRevA.75.013811}{Phys. Rev. A {\bf
  75}, 013811 (2007)}.

\bibitem{Scala2007JPA}
M.~Scala, B.~Militello, A.~Messina, S.~Maniscalco, J.~Piilo, and K.-A.
  Suominen, \href{http://stacks.iop.org/1751-8121/40/i=48/a=015}{J. Phys. A
  {\bf 40}, 14527 (2007)}.

\bibitem{Fleming2010JPA}
C.~Fleming, N.~I. Cummings, C.~Anastopoulos, and B.~L. Hu,
  \href{http://stacks.iop.org/1751-8121/43/i=40/a=405304}{J. Phys. A {\bf 43},
  405304 (2010)}.

\bibitem{Beaudoin2011PRA}
F.~Beaudoin, J.~M. Gambetta, and A.~Blais,
  \href{http://link.aps.org/doi/10.1103/PhysRevA.84.043832}{Phys. Rev. A {\bf
  84}, 043832 (2011)}.

\bibitem{Bamba2012DissipationUSC}
M.~Bamba and T.~Ogawa,
  \href{http://link.aps.org/doi/10.1103/PhysRevA.86.063831}{Phys. Rev. A
  {\bf 86}, 063831 (2012)}.


\bibitem{Blow1990PRA}
K.~J.~Blow, R.~Loudon, S.~J.~D.~Phoenix, and T.~J.~Shepherd,
  \href{http://link.aps.org/doi/10.1103/PhysRevA.42.4102}{Phys. Rev. A {\bf 42}, 4102 (1990)}.

\bibitem{Lang1973PRA}
R.~Lang, M.~O. Scully, and W.~E. Lamb,
  \href{http://link.aps.org/doi/10.1103/PhysRevA.7.1788}{Phys. Rev. A {\bf 7},
  1788 (1973)}.

\bibitem{Bamba2013MBC}
M.~Bamba and T.~Ogawa,
  \href{http://link.aps.org/doi/10.1103/PhysRevA.88.013814}{Phys. Rev. A
  {\bf 88}, 013814 (2013)}.

\bibitem{cohen-tannoudji89}
C.~Cohen-Tannoudji, J.~Dupont-Roc, and G.~Grynberg, {\em Photons and Atoms:
  Introduction to Quantum Electrodynamics},  (Wiley, New York, 1989).

\bibitem{Bamba2014SPT}
M.~Bamba and T.~Ogawa,
  \href{http://link.aps.org/doi/10.1103/PhysRevA.90.063825}{Phys. Rev. A
  {\bf 90}, 063825 (2014)}.

\bibitem{hopfield58}
J.~J. Hopfield, \href{http://dx.doi.org/10.1103/PhysRev.112.1555}{Phys. Rev.
  {\bf 112}, 1555 (1958)}.

\bibitem{Bamba2016Laser}
M.~Bamba and T.~Ogawa,
  \href{http://link.aps.org/doi/10.1103/PhysRevA.93.033811}{Phys. Rev. A
  {\bf 93}, 033811 (2016)}.

\bibitem{Mallory1969PR}
W.~R. Mallory, \href{http://link.aps.org/doi/10.1103/PhysRev.188.1976}{Phys.
  Rev. {\bf 188}, 1976 (1969)}.

\bibitem{Hepp1973AP}
K.~Hepp and E.~H. Lieb,
  \href{http://www.sciencedirect.com/science/article/pii/0003491673900390}{Ann.
  Phys. {\bf 76}, 360  (1973)}.

\bibitem{Wang1973PRA}
Y.~K. Wang and F.~T. Hioe,
  \href{http://link.aps.org/doi/10.1103/PhysRevA.7.831}{Phys. Rev. A {\bf 7},
  831 (1973)}.

\bibitem{Baumann2010N}
K.~Baumann, C.~Guerlin, F.~Brennecke, and T.~Esslinger,
  \href{http://dx.doi.org/10.1038/nature09009}{Nature {\bf 464}, 1301 (2010)}.

\bibitem{Baumann2011PRL}
K.~Baumann, R.~Mottl, F.~Brennecke, and T.~Esslinger,
  \href{http://link.aps.org/doi/10.1103/PhysRevLett.107.140402}{Phys. Rev.
  Lett. {\bf 107}, 140402 (2011)}.

\bibitem{Lolli2015PRL}
J.~Lolli, A.~Baksic, D.~Nagy, V.~E.~Manucharyan, and C.~Ciuti,
\href{http://link.aps.org/doi/10.1103/PhysRevLett.114.183601}
{Phys. Rev. Lett. {\bf 114}, 183601 (2015)}.

\bibitem{Bamba2016circuitSRPT}
M.~Bamba, K.~Inomata, and Y.~Nakamura,
\href{http://arxiv.org/abs/1605.01124}{arXiv:1605.01124 [quant-ph]}.


\end{thebibliography}

\end{document}